\RequirePackage{fix-cm}
\documentclass[smallextended]{svjour3}       % onecolumn (second format)
\smartqed  % flush right qed marks, e.g. at end of proof
\usepackage{graphicx}
\usepackage{mathptmx}      % use Times fonts if available on your TeX system
\usepackage{hyperref}
\usepackage{amsmath}
\usepackage{booktabs}
\usepackage{caption}
\usepackage{multirow}

\usepackage[title]{appendix}
\usepackage{subfigure}

\usepackage{soul}
\usepackage{color}
\definecolor{myhighlight}{rgb}{0.92,0.97,0.37}

\usepackage{xcolor,framed,graphicx}
\definecolor{shadecolor}{rgb}{0.92,0.97,0.37}

\journalname{}
\begin{document}

\title{Predicting the Objective and Priority of Issue Reports in Software Repositories}
\titlerunning{Predicting the Objective and Priority of Issue Reports}
\author{Maliheh Izadi, Kiana Akbari, Abbas Heydarnoori}
\authorrunning{M. Izadi, K. Akbari, A. Heydarnoori}

\institute{M. Izadi \at
              Automated Software Engineering research lab,\\
              Sharif University of Technology, Tehran, Iran.\\
              \email{maliheh.izadi@sharif.edu}\\
          \and
          K. Akbari \at
              \email{kakbari@ce.sharif.edu}\\          
          \and
          A. Heydarnoori \at
              \email{heydarnoori@sharif.edu}\\
}
\date{Received: date / Accepted: date}
\maketitle

\begin{abstract}
Software repositories such as GitHub
host a large number of software entities. 
Developers collaboratively discuss, implement, use, 
and share these entities. 
Proper documentation plays an important role 
in successful software management and maintenance.
Users exploit Issue Tracking Systems,
a facility of software repositories,
to keep track of issue reports,
to manage the workload and processes,
and finally, to document the highlight of their team's effort.
An issue report 
is a rich source of collaboratively-curate software knowledge,
and can contain 
a reported problem,
a request for new features, 
or merely a question about the software product.
As the number of these issues increases, 
it becomes harder to manage them manually.
GitHub provides labels for tagging issues, 
as a means of issue management.
However, about half of the issues 
in GitHub's top $1000$ repositories do not have any labels.

In this work, we aim at automating the process 
of managing issue reports for software teams.
We propose a two-stage approach to predict 
both the objective behind opening an issue 
and its priority level using feature engineering methods 
and state-of-the-art text classifiers. 
To the best of our knowledge, 
we are the first to fine-tune a Transformer for issue classification.
We train and evaluate our models 
in both project-based and cross-project settings.
The latter approach provides a generic prediction model
applicable for any unseen software project 
or projects with little historical data.
Our proposed approach can successfully 
predict the objective and priority level of issue reports 
with $82\%$ (fine-tuned RoBERTa) 
and $75\%$ (Random Forest) accuracy, respectively.
Moreover, we conducted human labeling and evaluation 
on unlabeled issues from six unseen GitHub projects 
to assess the performance of the cross-project model on new data.
The model achieves $90\%$ accuracy on the sample set. 
We measure inter-rater reliability 
and obtain average Percent Agreement of $85.3\%$ 
and Randolph's free-marginal Kappa of $0.71$ 
that translate to substantial agreement among labelers.

\end{abstract}

%%%%%%%%%%%%%%%%%%%%%%% Introduction %%%%%%%%%%%%%%%%%%%%%%%%%%%%%%%%%%%
%%%%%%%%%%%%%%%%%%%%%%%%%%%%%%%%%%%%%%%%%%%%%%%%%%%%%%%%%%%%%%%%%%%%%%

\section{Introduction}\label{sec:intro}
Due to the possibility of public discussions and contributions, 
Software Engineers and developers can collaboratively develop and maintain software projects.
In doing so, a growing base of knowledge has formed on software-related platforms such as GitHub and Stack Overflow.
This knowledge encapsulates various types of information 
such as source code, user reports, software Q\&A posts, and more. 
This raw yet invaluable knowledge can be transformed into 
automatic and practical solutions using data-driven approaches 
to help developers achieve their tasks more efficiently.

Most software repositories have a tracker
for recording and managing tasks of a project.
These trackers are the primary mean 
for communication, discussion, 
getting help, sharing opinions, making decisions, 
and finally collecting users' feedback.
GitHub's tracker is called \textit{Issues}.
Issue reports are an important source of knowledge provided with the help of the community. 
Any GitHub user is able to discuss, 
and contribute to the progress of a software project using issue reports.
Users can create an issue in a repository 
for various reasons including reporting bugs in the system, 
requesting new features, or asking for support.
This source of collaboratively-curated knowledge 
can be of great assistance in the process of software development and maintenance.
Team members should address these issues as soon as possible to keep their audience engaged 
and improve their software product.
As the project grows, the number of users and reported issues increases.
For instance, \textit{Elastic-search} project 
has more than $27K$ issue reports since $2017$.
It has on average, $25$ and $760$ daily and monthly new issues, respectively.
Consequently, timely management of issues 
including determining the goal of issues (classification of issue objectives), 
identifying urgent issues to address (prioritizing issues), 
and selecting the most important changes 
to include in product reports such as release notes, becomes harder.

\begin{figure}[tb!]
    \centering
    \includegraphics[width=\textwidth]{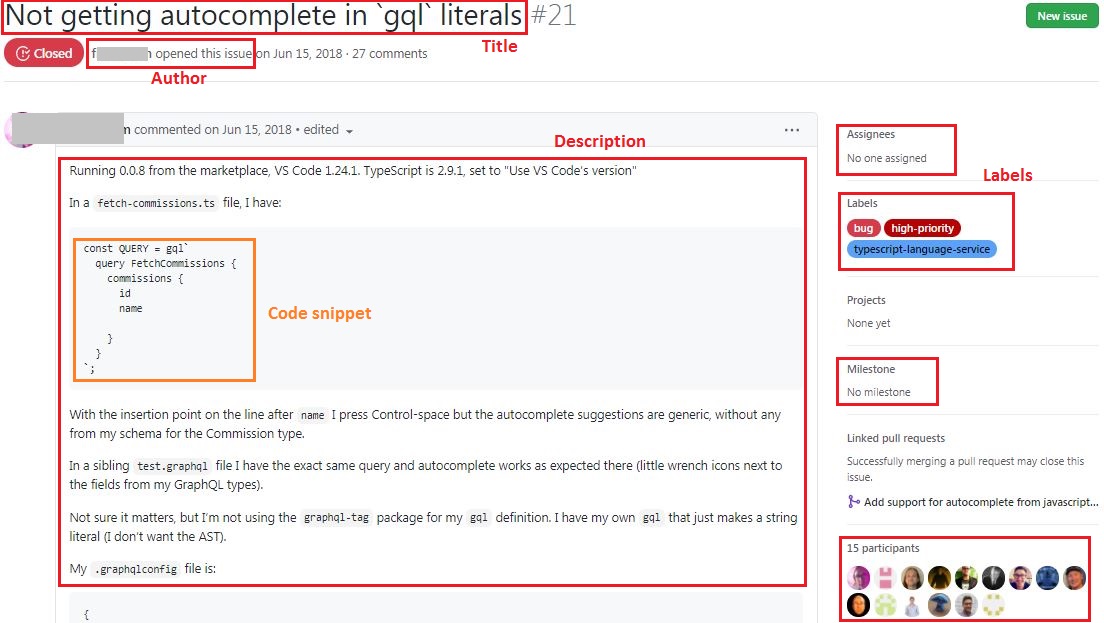}
    \caption{Issue sample}
    \label{fig:issue}
\end{figure}

Issues in software repositories 
must have a title, a description, and a state (open or closed).
They can also have additional data such as labels, 
assignee, milestone,  comments, etc.
Figure \ref{fig:issue} presents an issue from GitHub
which contains various types of information including 
title, description, author, and participants.
As shown, the description of this issue contains useful information 
including the reported problem and code snippets to elaborate the reported problem.
Moreover, it has several labels 
such as \texttt{bug report} to denote its objective 
and \texttt{high-priority} to indicate its importance.
Labels, as a sort of project metadata, describe the goal and content of an issue.
They are mainly used for categorizing, managing, searching, and retrieving issues.
Thus, assigning labels to issues
facilitates task assignment, maintenance and management of a software project.
Consequently, issue management is a vital part of the software development process.

Labels are assigned to issues
to indicate their objective, status, priority level, etc.
Such labels can help team members 
manage and track their tasks more efficiently.
Cabot et al. \cite{cabot2015exploring} 
analyzed about three million non-forked GitHub repositories  
to investigate the label usage and its impact on resolving issues. 
They showed only about $3\%$ of these repositories had labeled issues,
which indicates labeling issues is rarely done by developers.
Furthermore, in the repositories which incorporated issue labeling, 
only about $58\%$ of issues were labeled.
In their study, each issue had $1.14$ labels on average.
% They also found about $79\%$ of engaged users helped in resolving labeled issues.
The authors showed addressing an issue
and the engagement rate 
both have a high correlation with the number of labeled issues in a repository \cite{cabot2015exploring}.
% This indicates the act of labeling issues attracts more engagement 
% and results in faster resolving of issues.
This may indicate that labeling issues can benefit project management.
% Pham et al. \cite{cabot2015exploring} also claimed
% labeled issues are solved quicker than unlabeled issues.
Recently, Liao et al. \cite{liao2018exploring} 
investigated the effect of labeling issues on issue management.
They analyzed six popular projects 
and found labeled issues were addressed immediately, 
while unlabeled issues could remain open for a long time.
They also emphasized the need for correct labeling.
Previously, Herzig et al. \cite{herzig2013s} also
reported about $34\%$ of bug reports 
submitted by various users are misclassified (has a wrong label).
Misclassified reports can be misleading 
and result in a prolonged resolving process.
They can cause failed task assignment 
and/or impact the performance of bug prediction models.
This indicates the need for proper labeling of issue reports using an unbiased model.

In this study we consider two types of labels, 
namely \textit{objective} and \textit{priority} labels for an issue.
Based on our label analysis (refer to \autoref{sec:label_inspection}), 
we take the three most frequent reasons for opening issues as the main objectives.
These are \textit{Bug} reports, \textit{Enhancement} requests, and \textit{Support/Document}.
We also consider two priority levels, \textit{High} and \textit{Low}. 
The former should be addressed as soon as possible
while the latter can be handled with less urgency.
Detecting the priority level of issue reports has a two-fold gain;
not only it helps with 
accurate and timely resource allocation for bug triaging,
but also it results in less cost 
regarding maintenance and documentation purposes of the project.
For instance, the high-priority addressed issues 
can be listed in release notes or other performance reports of a project.

% models
Using a two-staged approach,
we aim to predict both the objective and priority of an issue.
We first predict an issue's objective by inspecting its textual information, 
namely its title and description.
We fine-tune a pretrained transformer-based model to classify issue objectives
into three categories of Bug, Enhancement, or Support/Document.
More specifically, 
we adapt the the Robustly-optimized BERT approach (RoBERTa)~\mbox{\cite{liu2019roberta}} 
proposed by Facebook to our case.
Our experiments
indicate that using these types of textual information 
is sufficient for successfully predicting these objectives.
In the second stage, to train our classifiers 
we define three sets of features, 
namely Textual Features, 
Label Features, 
and Normalized Features 
that can potentially help in predicting the importance of an issue.
Textual Features include TF-IDF vectors of title and description of issues.
Label Features are one-hot encoded vectors of available labels for an issue.
For the third input vector, Normalized Features, 
we apply feature engineering methods
and scale the numerical information
from five different information resources
including textual-based, developer-related, discussion-related, 
event-related, and sentiment of the issues.
Finally, we train multiple classifiers for predicting the priority of an issue.
We obtain the best result using a Random Forest (RF) classifier.

% dataset and results
For the first task, 
we use about $817,743$ issues and train a single generic model applicable for all repositories.
For the second task, 
we train our models in both project-based and cross-project settings 
using about $82,719$ issues.
We evaluate our models in both tasks using standard metrics 
including Precision, Recall, F1-measure, and Accuracy. 
Our fined-tuned RoBERTa-based classifier achieves $82\%$ of accuracy, 
outperforming baseline models.
Moreover, our priority prediction model scores $75\%$ of accuracy. 
The results show that both project-based 
and cross-project prediction models 
for the second task perform comparably.
Therefore, our model is expected to efficiently work for unseen repositories 
without the need for more training.
Nonetheless, we conducted a human labeling and evaluation experiment
to assess the proposed model's performance on new data, 
i.e.,  unlabeled issue reports from six unseen projects.
Sixty issues were randomly selected from these projects. 
Thirty Software Engineers participated in our study, 
and we collected $300$ votes for the sample set.
The results indicate the high accuracy of the proposed model on unseen data. 
Moreover, we also asked the participants the factors they take into account 
while determining priority levels of issues and report their insights in this work.
Our contributions are:
\begin{itemize}
    \item We train a model to predict issue objectives (bug report, enhancement, and support) 
    and obtain $82\%$ accuracy. 
    To the best of our knowledge, we are the first to 
    adapt transformer-based models to predict labels for issue reports.
    \item We train project-based models for predicting the priority 
    of issue reports using feature engineering methods and state-of-the-art text classifiers.
    We also train a generic model for priority prediction 
    in a cross-project setting. This model performs on par with the project-based models with $74\%$ accuracy.
    \item We conducted a human labeling and evaluation task
    to assess the performance of the proposed model on unseen data and achieved high accuracy ($90\%$).
    We obtain Percent Agreement of $85.3\%$ and Kappa of $0.71$ 
    which translate to \textit{substantial} agreement among our participants.
    \item We collected and preprocessed two sets of large-scale datasets with objective and priority labels from GitHub. We manually inspected synonym but differently-written labels and clustered them to decrease noises in user-defined tags.
    We release our source code and datasets for replication and use by other researchers.\footnote{\url{https://github.com/MalihehIzadi/IssueReportsManagement}}\footnote{\url{https://zenodo.org/record/4925855\#.YNME2r4zbtQ}}
\end{itemize}

%%%%%%%%%%%%%%%%%%%%%%%%%%%%%%%%%%%%%%%%%%%%%%%%%%%%%%%%%%%%%%%%%%%%%%%%%%%%
%%%%%%%%%%%%%%%%%%%%%%%% Approach %%%%%%%%%%%%%%%%%%%%%%%%%%%%%%%%%%%%%%%%%%
%%%%%%%%%%%%%%%%%%%%%%%%%%%%%%%%%%%%%%%%%%%%%%%%%%%%%%%%%%%%%%%%%%%%%%%%%%%%

\section{Approach}\label{sec:approach}
In this section, 
we first present an overview of our proposed approach.
Then, we elaborate on each phase with more details.

\subsection{Approach Overview}
\begin{figure}[ht!]
    \centering
    \includegraphics[width=\textwidth]{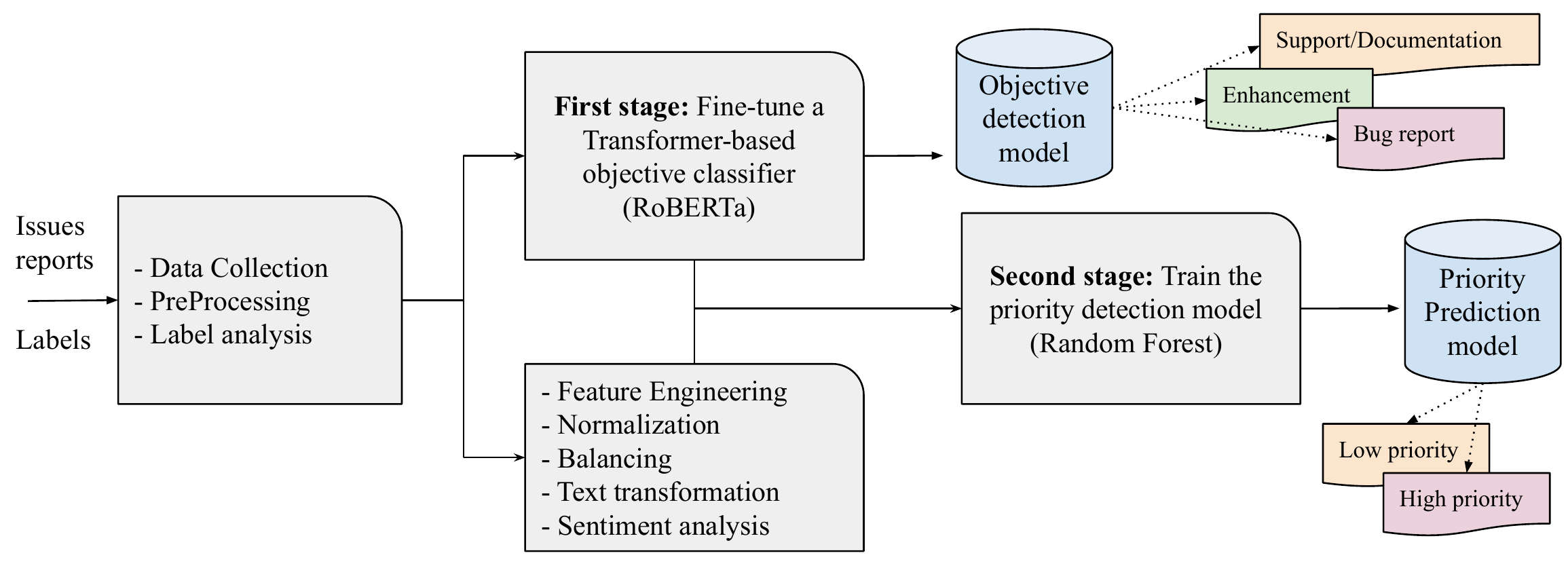}
    \caption{Summary of the proposed approach}
    \label{fig:approach_summary}
\end{figure}

Figure~\mbox{\ref{fig:approach_summary}} 
presents a concise summary of our proposed approach.
Our two-stage approach for predicting the objective and priority of issues consists of
(1) analyzing issue labels on GitHub to determine which labels to use in our training,
(2) data collection and preprocessing,
(3) issue-objective prediction, 
(4) feature engineering and model training,
and finally (5) predicting priority labels.

We first collect the data of issue reports 
using the GitHub API.\footnote{\url{https://developer.github.com/v3/}}
Then, we extract textual information of issue reports, 
i.e., their title and description.
We also extract all labels assigned to issues.
Finally, we process and save $73$ types of information from these reports 
(such as the author, closer, events, milestones, comments, etc.).
Then we perform rigorous text processing techniques on the data.

In the next phase, 
we train a transformer-based classifier, 
to predict the objective of an issue.
More specifically, 
we fine-tune RoBERTa~\mbox{\cite{liu2019roberta}} on our dataset.
The three intended categories we use are Bug Report, Enhancement, and Support/Documentation.

\begin{figure}[tb!]
    \centering
    \includegraphics[width=0.9\textwidth]{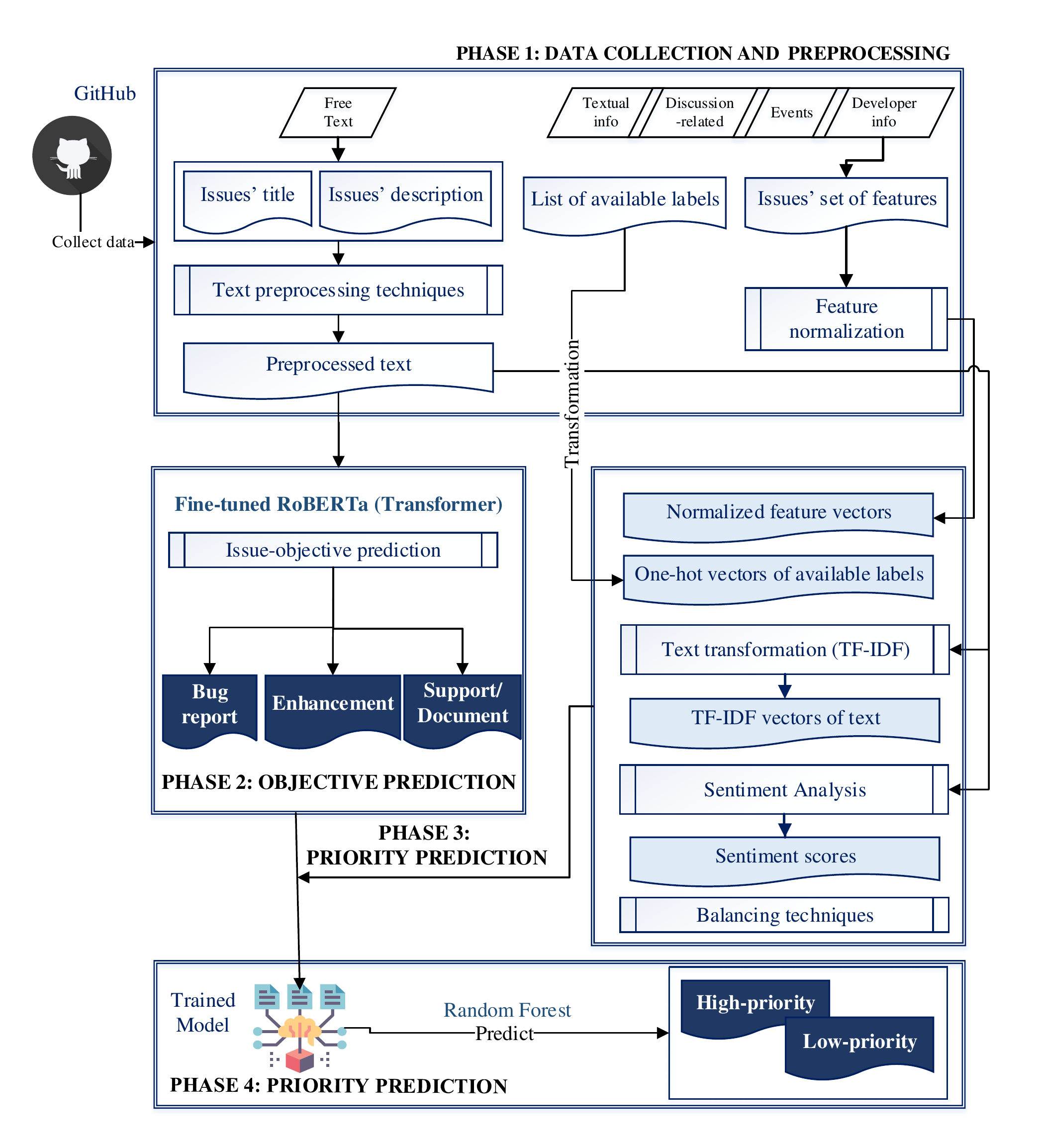}
    \caption{Approach workflow}
    \label{fig:approach}
\end{figure}

In the third phase, we take the information 
we gathered in the previous phases 
and employ various NLP and Machine Learning techniques 
to train a model based on RF
for predicting priority levels of issues.
Finally, we use our cross-project trained model 
to predict the priority of issues in unseen repositories.
More specifically, we conducted an experiment for human labeling and evaluation
to assess the performance of 
the proposed model 
on unlabeled issues 
from six unseen GitHub projects.
In the following sections, 
we provide more details for each step of the proposed approach.
Figure~\mbox{\ref{fig:approach}} presents the workflow of our proposed approach with more details.

\subsection{Label Inspection}\label{sec:label_inspection}
GitHub has a set of seven default issue labels, namely 
\texttt{bug}, \texttt{enhancement},  \texttt{question},
\texttt{help-wanted}, \texttt{duplicated}, \texttt{wont-fix}, and \texttt{invalid} . 
Members can also add or modify labels to suit their project's needs.

\begin{figure}[tb!]
    \centering
    \includegraphics[width=0.7\textwidth]{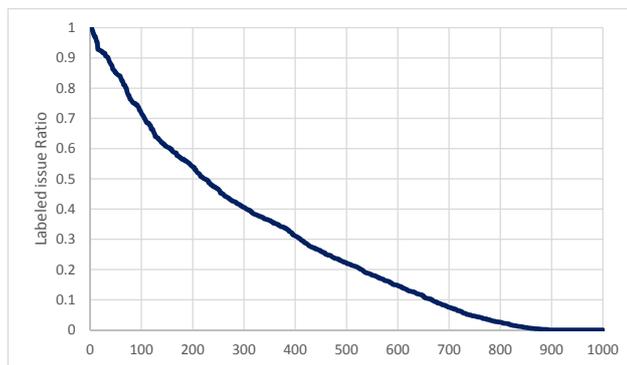}
    \caption{Labeled issues ratio per repository (for the top 1000 GitHub repositories)}
    \label{fig:labels_usage_per_repo}
\end{figure}

% our analysis
To obtain a better understanding of which labels we should use for each task 
(objective and priority prediction),
we collected labels used in the top $1000$ repositories of GitHub which had at least $500$ stars
using the GitHub API.\footnote{\url{https://api.github.com/search/repositories?q=stars:>500&sort=stars}}
These repositories are ranked based on their number of stars.
Then two of the authors analyzed the labels.
At the time we collected labels of these repositories,
they had $4,888,560$ issue reports in total, from which $2,333,333$
had at least one label.
This means approximately half of the issues of popular repositories did not have any labels.
Furthermore, on average, $71\%$ of all issue reports in each repository were unlabeled.
As shown by Figure \ref{fig:labels_usage_per_repo},
only $3\%$ of these repositories have labeled most of their issues (above $90\%$ coverage),
while about $80\%$ of repositories have labeled less than half of their issues.

\begin{figure}[tb!]
    \centering
    \includegraphics[width=\textwidth]{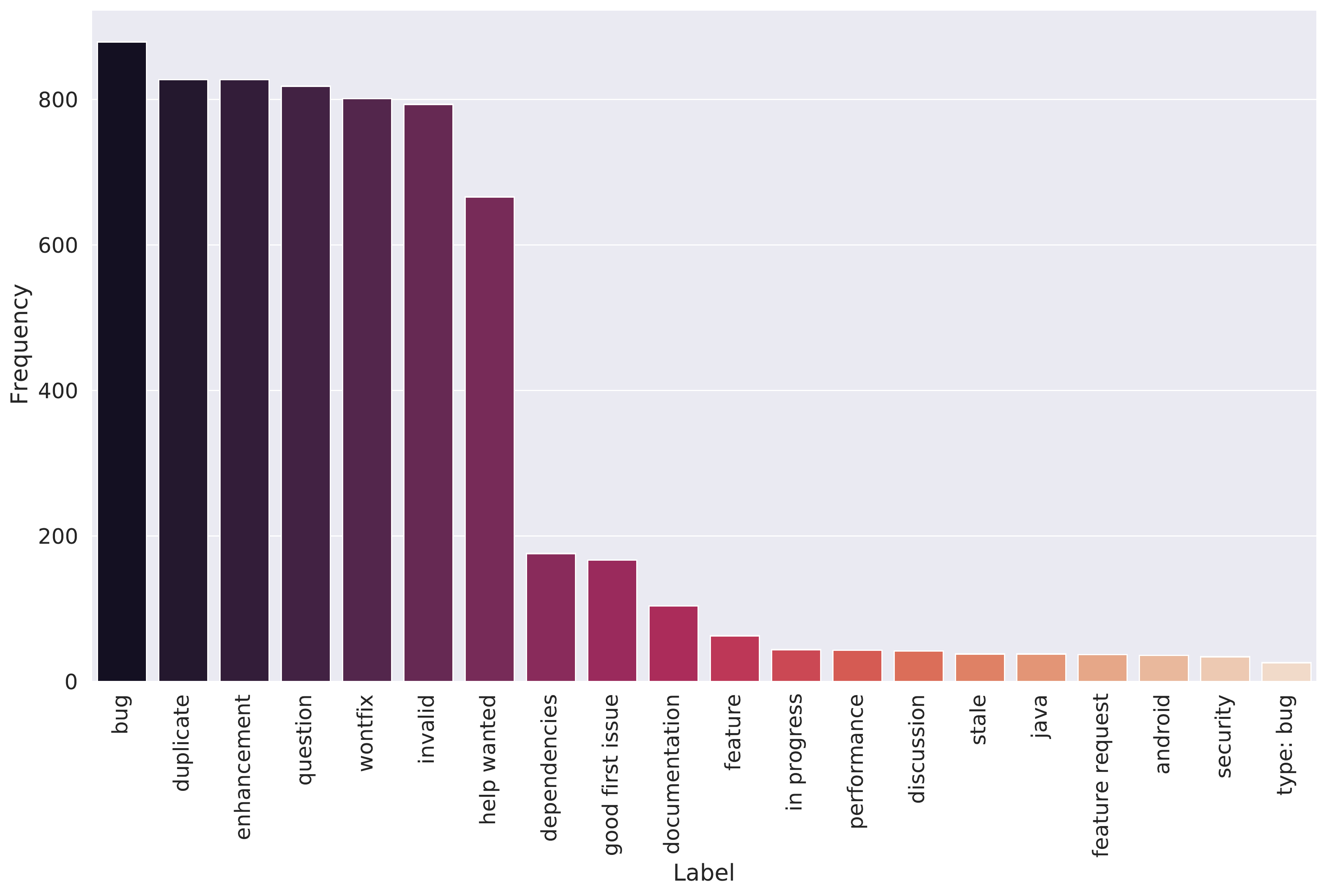}
    \caption{Label frequency among the top $1000$ GitHub repositories}
    \label{fig:labels_freq}
\end{figure}

Figure \ref{fig:labels_freq} shows top $20$ labels 
used in the most popular repositories of GitHub.
As expected, most of these repositories 
already have the above-mentioned seven default labels of GitHub.
So the frequency of these labels 
are much higher than the new customized labels defined by users.
We found $6182$ distinct labels in the top $1000$ repositories.
As shown, the frequency distribution has a long-tail.
However, labels, like any other tag entity, are written in free-format.
Thus, the distributed nature of the tagging process results 
in multiple differently-written labels with a common semantic.
For instance, issues opened to report bugs are tagged with labels such as 
\texttt{bug} and \texttt{type: bug} 
or issues for requesting new features 
are tagged with labels such as 
\texttt{feature}, \texttt{feature request}, and \texttt{new feature}.

Previous studies have also investigated the main categories for issue objectives~\cite{fan2017road,kallis2019ticket,bissyande2013got,cabot2015exploring}.
Upon investigating issues of three million repositories,
Cabot et al.~\cite{cabot2015exploring} concluded 
the most frequent issue labels in GitHub are  
\texttt{enhancement}, \texttt{bug}, \texttt{question}, 
\texttt{feature}, \texttt{documentation}, \texttt{won't fix}, and \texttt{task}. 
In another large-scale study on issue reports,
Bissyande et al.~\cite{bissyande2013got} 
analyzed about $800K$ issues from which $27\%$ were labeled.
They reported that the most frequent labels in their study 
were \texttt{bug} and \texttt{feature}.
Fan et al.~\cite{fan2017road} conducted a study 
to determine whether issue reports are related to bugs or not. 
They used the dataset provided by Yu et al.~\cite{yu2015wait} 
which contained $952$K issue reports from $1,185$ GitHub repositories. 
Among the $7,793$ labels in the dataset, $149$ were identified 
as the labels which indicate the \textit{type} of an issue. 
Over $252$K issue reports ($26\%$) in the dataset were tagged with one of these type labels. 
Fan et al.~\cite{fan2017road} categorized
the most frequently-used type labels 
into two major classes of 
\textit{bug-related} ($52\%$) and \textit{non-bug related} ($38\%$). 
The latter consists of the following labels: 
enhancement, feature, question, feature request, documentation, improvement, and docs.
This category can be broken down to two finer categories of \textit{Enhancement} and \textit{Support/Documentation}. 
Lastly, Kallis et al.~\cite{kallis2019ticket} 
also categorized issue reports into three classes of \texttt{bug}, \texttt{feature}, and \texttt{question}. 
Therefore, based on our analysis and previous studies, 
we selected the three most-frequently-used labels 
for issues' objectives in the top projects as
\textit{Bug Report}, \textit{Enhancement} and \textit{Support/Documentation}. 
Next, two of the authors independently and manually identified the most related 
but differently-written user-defined labels 
as these three main objectives. 
In this process, authors have relied on the definitions provided by GitHub for labels.\footnote{https://docs.github.com/en/issues/using-labels-and-milestones-to-track-work/managing-labels}
Then the authors compared the categories and discussed any conflicts to validate the final decision.
As a result, we collected issue reports 
that had at least one of labels mentioned in~\autoref{tab:label_obj} 
for each objective category.
Note that we only use mono-labeled issues in our dataset. 
Thus issues tagged with more than one label are removed.

% other labels
Note that there are other objectives for opening an issue, 
e.g., for testing, making announcements, or discussing matters in the team.
However, there were less frequently used compared to our main selected categories.
Moreover, among the most frequent labels, 
there are also other recurrent labels such as
\texttt{duplicate}, \texttt{wont fix}, \texttt{invalid}, 
\texttt{in progress}, \texttt{good first issue}, 
\texttt{stale}, \texttt{java}, \texttt{android}, etc.
However, these labels do not address the reason behind opening an issue.
They are merely other types of metadata for adding extra information.
That is why we do not include these labels as issue objective.
%%%%%
\begin{table}[tb!]
\caption{Selected labels for each category of objective}
\label{tab:label_obj}
\begin{tabular}{p{20mm}p{90mm}}
\toprule
\textbf{Category} & \textbf{Labels' list} \\
\toprule
Bug report &  
bug, defect, kind/bug, type: bug\\\midrule
Enhancement & 
enhancement, kind/enhancement, type: enhancement, \newline
type: improvement, improvement, \newline
feature request, feature, kind/feature, type: new feature, new feature \\\midrule
Support/ \newline Documentation & 
help wanted, status: help wanted, type: support, supports, \newline
question, type: question, kind/question, \newline
docs, documentation, type: documentation, kind/documentation, \newline
information, more info needed, more info required, more-information-needed, need more info, needs info, needs more info, needs-info, needs-details\\
\bottomrule
\end{tabular}
\end{table}

The second task which is prioritizing issues, 
requires issues with a priority-related label.
Therefore, we inspected various priority labels including
\texttt{blocker-priority}, \texttt{critical-priority}, \texttt{high-priority}, 
and \texttt{low-priority}.
Note that priority labels are also written in different formats.
For instance, we found the following labels 
as indicators of an issue with \textit{critical-priority}:
\texttt{criticalpriority}, \texttt{priority-critical},
\texttt{critical priority},	\texttt{priority:critical},	
\texttt{priority critical}, \texttt{priority: critical},	
\texttt{priority - critical}, \texttt{critical-priority}, 
\texttt{priority/urgent}, \texttt{priority/critical},
\texttt{critical}, and \texttt{urgent}.
Thus, to find these semantically similar labels, 
we performed the same analysis on priority-related labels explained above,
and found semantically similar but differently-written priority labels.

% Although the number of distinct user-specified labels are high, 
% considering the semantic similarity of some of these labels,
% we believe it is more useful 
% to only consider the most important labels for managing issues.

Moreover, in Section \ref{sec:label_clusters}, 
we report the result of our analysis 
for extracting other frequent and semantically-similar labels 
and incorporating them in our proposed approach 
as an optional feature vector.

\subsection{Data Collection}\label{sec:data_collection}
For the first task of predicting objective of an issue,
we collected closed issues from GitHub's open-source repositories 
with Java as their main programming language 
which were created before April $2021$.
We used Java as it was used frequently in previous studies 
and also to limit the number of retrieved issues.
We selected three main categories of objectives, namely
\textit{Bug Report}, \textit{Enhancement}, and \textit{Support/Documentation}
based on the labels presented in \autoref{tab:label_obj}.
The initial dataset for the classifying issue objectives 
contained $1,096,704$ issues from $79,729$ repositories.
Issues are grouped into three categories;
$480K$ bug-related issues, 
$528K$ enhancement-related issues, 
and $173K$ support-related issues.
In the end, after performing all our preprocessing steps reviewed below, 
there remained $817,743$ issues from $60,958$ repositories.
More specifically, we include
$362K$, $342K$, and $112K$ preprocessed issues belonging
to the bug report, enhancement, and support/documentation categories, respectively.
We denote this dataset as the \textit{issue-objective dataset}.

For the second task we collected issues 
with at least one of the following four priority-related labels:
\texttt{blocker-priority}, \texttt{critical-priority}, \texttt{high-priority}, 
and \texttt{low-priority}.
We aggregate issues with labels of 
blocker, critical and high-priority 
in the same group of the crucial issues.
The rest are categorized in the low-priority group.
In all, we collected $47$ synonyms for the two categories of High 
and Low priority 
(for the complete list refer to Appendix~\ref{appendix:priority_labls}.)
In the end, after preprocessing the data, 
we have a dataset of $82,719$ issues from $70$ repositories for this task. 
The preprocessed dataset contains $44,733$ \texttt{high-priority} 
and $37,986$ \texttt{low-priority} issues.
We denote this dataset as the \textit{Issue-priority dataset}.

\begin{figure}[tb!]
    \centering
    \subfigure[Objective categories]
    {\includegraphics[width=0.4\linewidth]{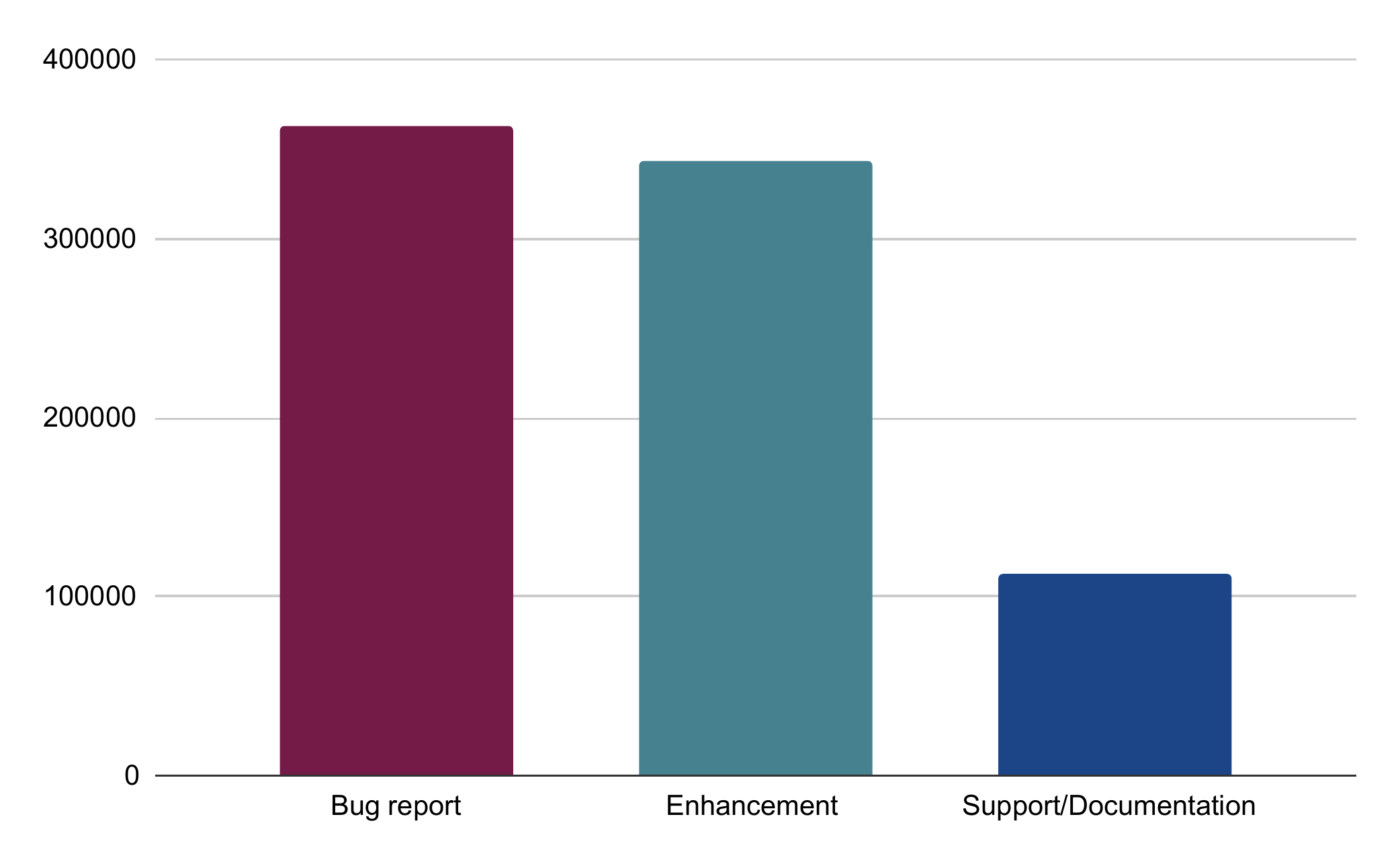}}
        \label{fig:obj_dist}
    \subfigure[Priority categories]
    {\includegraphics[width=0.4\linewidth]{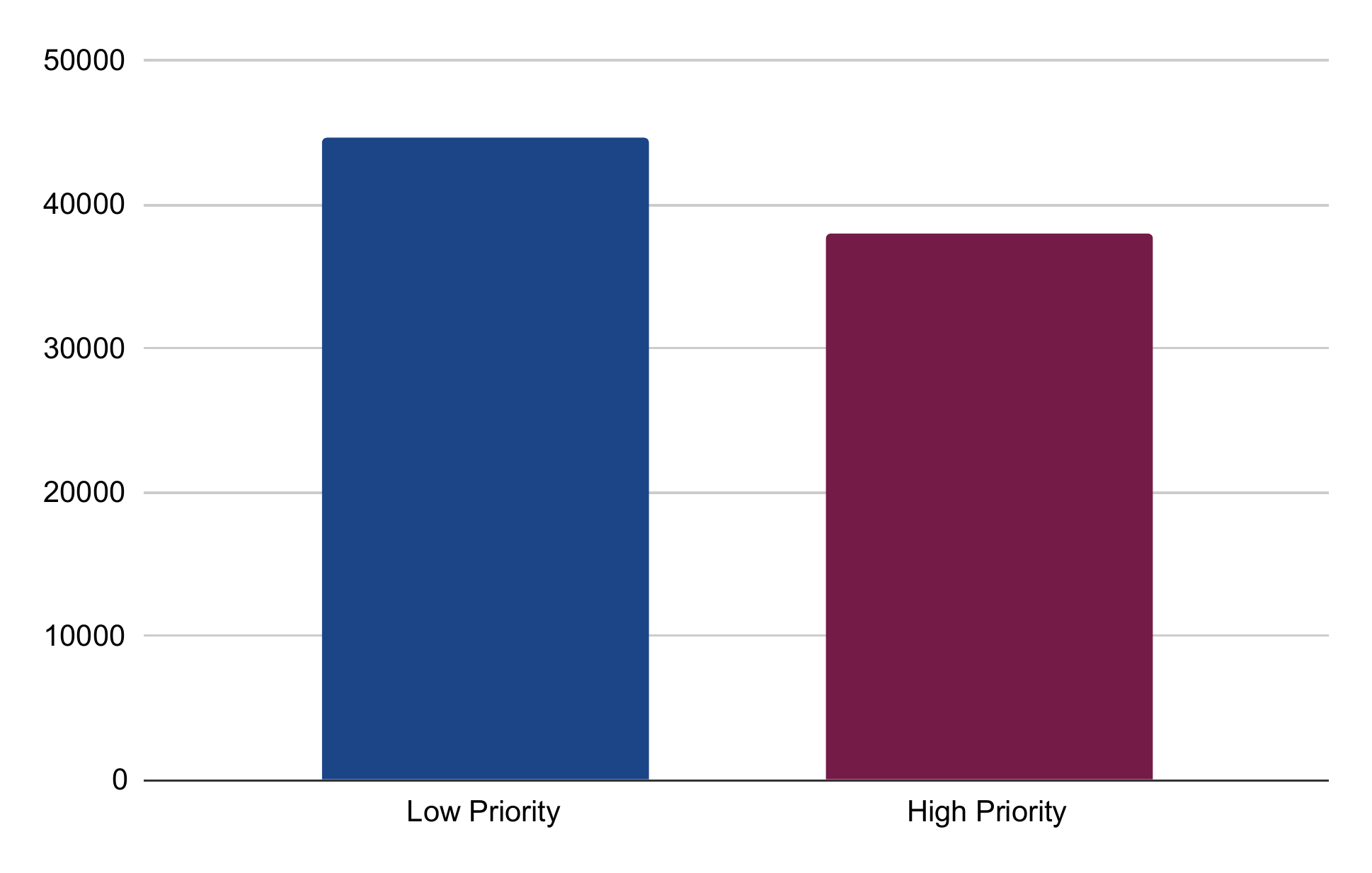}}
        \label{fig:prio_dist_all}
    \caption{Class distribution}
    \label{fig:class_dist}
\end{figure}

\begin{figure}[tb!]
    \centering
    \includegraphics[width=0.8\textwidth]{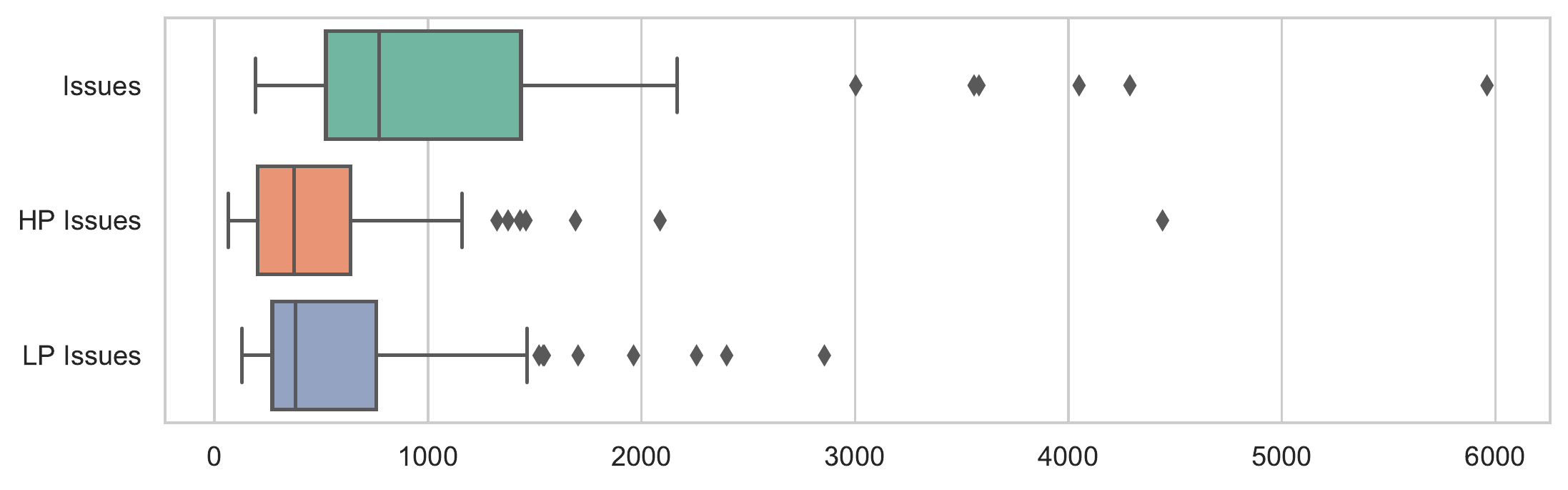}
    \caption{Distribution of LP and HP issues among repositories}
    \label{fig:issue_num}
\end{figure}

\begin{figure}[tb!]
    \centering
    \includegraphics[width=0.6\textwidth]{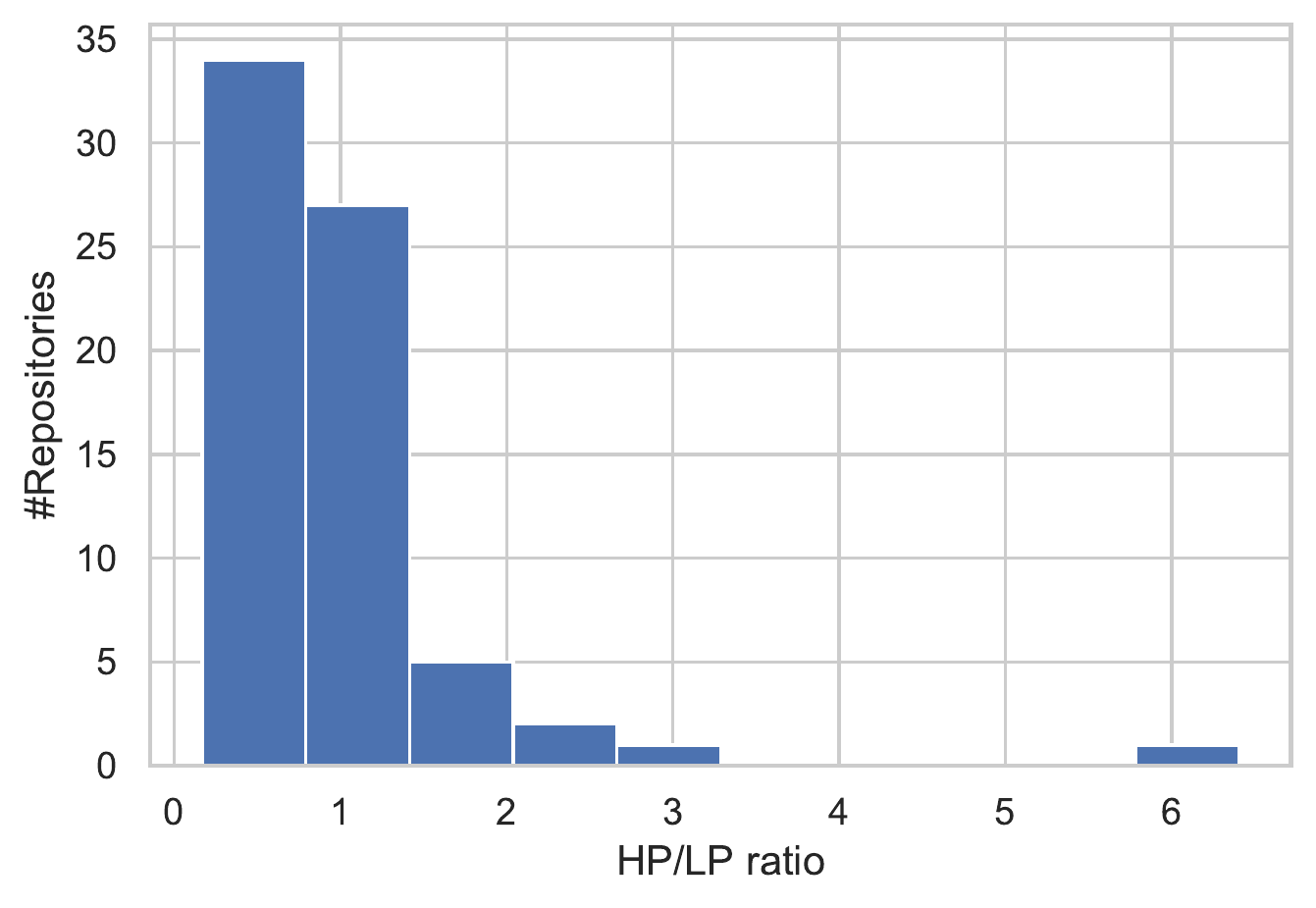}
    \caption{Histogram of HP/LP ratio per project}
    \label{fig:prio_hist}
\end{figure}

\autoref{fig:class_dist}
provides the distribution of objective and priority classes after preprocessing
in their respective dataset for all projects.
\autoref{fig:issue_num} presents three box-plots 
for the number of issues in the $70$ repositories of the Issue-priority dataset.
HP Issues and LP Issues 
denote the number of issues with High-Priority and Low-Priority labels, respectively.
\autoref{fig:prio_hist} depicts the ratio of 
HP to LP labels of issue reports ($HP/LP$) per projects in this datast.
Although the average $HP/LP$ ratio is $1.00$, 
this ratio per project ranges from $0.16$ to $6.40$. 
That is for some repositories, the HP class is more represented, 
while for others, the LP class is more supported. 
%%%%%%
%%%

\subsection{Preprocessing}
Each issue has two main textual information sources, 
namely title and description. 
To train our models,
we create a feature vector for both of them based on the following
preprocessing steps.

\textbf{Filtering and Cleaning}:
We first remove issues that have 
very little (less than three characters) or no text 
in their title or description.
We also remove issues that are tagged as \texttt{not an issue} 
or \textit{duplicate} issue reports to prevent biasing our models.
Then, we filter out issue reports 
that are written in a non-English language (more than $50\%$ of the text).
Then, we clean issues' textual information
by removing arbitrary digits, non-ASCII characters, and punctuation marks.
Note that we retain question marks 
as they are mainly used in questions and support related issues.
Thus, they can be helpful for predicting this class.

\textbf{Text Normalization}:
Handling large vocabularies is a challenging task in NLP-based researches.
Generally, studies limit vocabulary to the most common words 
and replace out-of-vocabulary tokens 
with a special unknown token \texttt{<UNK>}.
To reduce out-of-vocabulary tokens,
we normalize issue reports' textual information
using several normalization rules.
More specifically, we replace \textit{abstract concepts} such as 
\textit{user-names}, \textit{code snippets}, \textit{function calls},
\textit{markdown} symbols, \textit{emails}, \textit{URLs},  \textit{paths},
\textit{dates}, and \textit{times} using regular expressions.
The intuition is that including the exact content of these concepts 
increases the size of our vocabulary,
however, by performing text normalization 
we can both keep the notion of having e.g., a code snippet in an issue 
and remove the exact characters of that code snippet
to help our models learn better.
To achieve this, we replace the content of a code snippet 
with an abstract token \texttt{<CODE>}. 
We apply the same technique to the rest of the above-mentioned concepts as well.
Text normalization has been used before 
in preprocessing data for Machine Learning models \cite{svyatkovskiy2020intellicode,izadi2021topic}.

\textbf{Tokenization and Lemmatization}:
We split tokens based on several naming conventions 
including SnakeCase, camelCase, and underscores 
using an identifier splitting tool called 
\textit{Spiral}.\footnote{\url{https://github.com/casics/spiral.}} 
This will also mitigate the out-of-vocabulary problem.
Using NLTK library \footnote{\url{https://www.nltk.org/}}, 
we first tokenize the text of issue reports,
then we remove frequently used words in the language called stop-words 
which do not bring any value to the models. 
Note that we keep negative words such as NOT 
and compulsory words such as MUST, 
which can be useful for the sentiment analysis phase.
We then lemmatize the preprocessed text 
to reduce grammatical forms
but retain their correct word formats.

\textbf{Transformation}:
The final step is to transform 
the textual information of issues
to their mathematical representation 
that can be fed to the Machine Learning models.
We convert the collection of preprocessed issues' text
to a matrix of TF-IDF vectors.
More specifically, we represent each issue title and description 
as a vector where each word is a feature. 
% We use the Term Frequency (TF) of a word divided by 
% Document Frequency (IDF) to get the value of each word feature.
Note that we generate their TF-IDF embedding vectors separately, 
then we concatenate these two vectors for each issue.
The simpler approach would be to first 
concatenate the text of these two sources 
and then build the embedding vectors. 
However, our experiment yielded better results 
with separate title and description TF-IDF vectors.
This is probably due to the fact 
that although both title and description 
are inherently textual information describing an issue,
their abstract level and objective differ. 
Interestingly, our Machine Learning models 
were capable of picking up on this difference.
It is worth mentioning we also experimented 
with Doc2Vec and Word2Vec embeddings. 
However, TF-IDF vectors yielded the best results,
thus we only report them in this work.

\subsection{First Stage: Objective Detection}\label{sec:model_objective}
Previously, we obtained and preprocessed our issue-objective dataset.
In this step, we train our classifier for the first stage of our approach.
To predict issue objectives (Bug report, Enhancement, or Support/Document), 
we train a transformer-based model on the issue-objective dataset.
We fine-tune the RoBERTa~\cite{liu2019roberta} on our issue-objective dataset.
RoBERTa includes pretraining improvements 
(compared to the vanilla BERT model \cite{devlin2018bert})
using only unlabeled text from the web, 
with minimal fine-tuning and no data augmentation. 
The authors modified the masked language modeling task of BERT by using dynamic masking 
based on a new masking pattern generated each time a sentence is fed into training. 
They also eliminated the next sentence prediction task since Facebook's analysis indicated 
that it indeed hurts the model's performance.
The Input of our model is the concatenated preprocessed 
word vectors of title and description of issues.
In this stage, 
we feed the models with preprocessed text (word vectors) 
and the models process them accordingly.
The output of the first stage 
is the probability of an issue to be a Bug report, Enhancement, or Support.

\subsection{Sentiment Analysis}
Sentiment analysis uses computational linguistics and NLP techniques 
to quantify the intended sentiment of a piece of text.
We believe more urgent issues hold more distinct sentiment, 
and use sentiment analysis methods to extract this information 
from textual information of issues.
We use \textit{SentiStrength} which quantifies the strength of positive and negative sentiment\footnote{\url{http://sentistrength.wlv.ac.uk/}} in text. 
SentiStrength reports two scores 
in the range of ($-1$, $-5$) with $-5$ for extremely negative sentiment, 
and ($1$, $5$) with $5$ as the extremely positive sentiment.
Psychology research claims that we process positive and negative sentiment at the same time.
Thus, SentiStrength reports both sentiment scores (positivity and negativity).
We apply SentiStrength on both issues' title and description 
and analyze these features in our feature selection process.

We also use TextBlob, a library that quantifies sentiment 
in terms of two measures of subjectivity and polarity.\footnote{\url{https://textblob.readthedocs.io/en/dev/}}
It reports a tuple of $Sentiment(polarity, subjectivity)$. 
Polarity range is $[-1, 1]$ and subjectivity range is $[0, 1]$ 
with $0$ as completely objective and $1$ as completely subjective.
Same as the above, we use this library on the title and description of issues 
and use them in the feature selection process.

\subsection{Label Clustering}\label{sec:label_clusters}
Labels are free-format text.
Thus, users can use different word formats
for semantically-similar concepts.
Clustering the morphological synonym labels
as a form of issue label management
can  boost the performance of Machine Learning models 
which takes these labels as inputs. 
However as shown in section \ref{sec:label_inspection}, 
the number of distinct user-specified labels is high.
To be able to decrease issue labels' space, 
two of the authors manually analyzed the collected labels from the top 1000 repositories, 
and found several clusters of semantically similar labels.
Based on our investigation, 
we selected the most $66$ frequently-used labels in GitHub 
and then extracted their synonyms but differently-written labels
to build a dataset of clusters of labels.
\autoref{tab:label_clusters} shows two sample identified 
clusters.\footnote{A complete list of these $66$ clusters is available in our repository.}
%%%%
\begin{table}[tb!]
\caption{Semantically similar clusters of issue labels}
\label{tab:label_clusters}
\begin{tabular}{lp{95mm}}
\hline\noalign{\smallskip}
Representative & Semantically similar labels \\
\noalign{\smallskip}\hline\noalign{\smallskip}
Duplicate & duplicate, status/duplicate, status: duplicate, status:duplicate, status=duplicate, status-duplicate, type:duplicate, was:duplicate, resolution:duplicate, resolution/duplicate, duplicate issue, t-duplicate, r: duplicate, closed: duplicate, kind/duplicate, type: duplicate \\
Won't fix & won't fix, wont fix, wontfix, wont-fix, status: won't fix, will not fix, resolution:won't fix, status=will-not-fix, closed: won't fix, state:wont-fix, status: will not fix, won't-fix, will-not-fix, cant-fix, cantfix, can't fix \\
\noalign{\smallskip}\hline
\end{tabular}
\end{table}

We use representatives of these clusters as one of our feature sets.
More specifically, we build a one-hot vector with a size of $66$, 
in which each element denotes the presence of one of the label clusters.
In the model construction phase, 
we concatenate this label vector 
with the TF-IDF embedding vector of textual information of an issue 
and the selected features' normalized vector 
and feed the final vector to our model.

\subsection{Feature Extraction and Categorization}
Before training our prioritizer classifier, 
we performed feature selection.
Feature selection is the process of 
selecting a subset of relevant predictors to feed to the Machine Learning model.
These techniques are usually employed 
to simplify models,
provide better interpretation,
avoid overfitting by providing more generalizable models,
and achieve a shorter training time \cite{dash1997feature}.

Two of the authors manually inspected issue reports 
and extracted a list of $73$ potential features 
which could affect the importance of issue reports.
These features included (but are not limited to) information about 
textual length of an issue, 
author of an issue, 
the closer of an issue,
were the author and closer the same people,
the amount of discussion an issue has attracted, 
how long the discussions took,
the sentiment of the discussions,
number of events on the issue, 
does it have a commit, milestone, or assignee,
is it a Pull Request,
and many more.
Each of these features can potentially affect the outcome.
For instance, experienced developers 
are more likely to report or close
important issues.
A heated and/or long discussion can be 
an indication of urgent matters being discussed by the team members.
For each opened Pull Requests, an issue is opened automatically~\cite{kalliamvakou2014promises}. 
Pull Requests can be considered as important issues. 
In fact using GitHub API when collecting the dataset, 
one can choose to retrieve only issues 
and exclude Pull Requests or retrieve all of them together.  
Considering the tight relationship of issues and Pull Requests, 
we decided to keep Pull Requests when collecting the data. 
Note that Pull Requests can also be investigated separately using their specific features and applications~\cite{gousios2015work,veen2015automatically}.
Therefore, we compute the correlation of these features, 
draw their heat map (filter-based selection),
perform two wrapper-based selection methods, 
namely backward and recursive feature elimination approaches 
to analyze these features and remove rudimentary ones.
In the end, we choose $28$ features and categorize them into five groups of 
\textit{textual-}, \textit{discussion-}, 
\textit{events-}, \textit{developer-}, 
and \textit{sentiment-related} features.
Table \ref{tab:feature_selection} summarizes these features.
Our analysis showed 
text length and the existence of code snippets and URLs inside the description can help the model.
For the discussion-related features, 
we include four features, namely
number of comments, the average length of comments, 
the ratio of the number of comments to the number of engaged developers in the discussion,
and discussion time.
For the events-related category, we include six features: 
the number of all events,
the fact that whether this issue is assigned, 
does it have a milestone already, 
is it a Pull Request,
does any commit reference this issue,
and finally, how many labels does it have.
For the developer-related category, we use ten features 
including various information about who has opened the issue, 
their reputation and number of followers/followings,
their experience and contribution to this project and GitHub in general,
their association, 
that is, whether they are a team member or merely a GitHub user,
their profile age and whether the author and closer are the same users or not.
Because the author and closer information 
have a high correlation score (above $80\%$) in our dataset,
we do not include \textit{closer information} separately.
%%%
\begin{table}[tb!]
\caption{Selected features for prioritizing issue reports}
\label{tab:feature_selection}
\begin{tabular}{llp{70mm}}
\hline\noalign{\smallskip}
Category & Feature & Description  \\
\noalign{\smallskip}\hline\noalign{\smallskip}
\multirow{4}{*}{Textual information} & title-words & Number of words in the title of an issue. \\
                                     & desc-words  & Number of words in the description of an issue \\
                                     & code    & Number of code snippets in the description of an issue \\
                                     & url     & Number of URLs in the description of an issue \\\hline

\multirow{4}{*}{Discussion} & comments & Number of comments in the discussion \\
 & cm-mean-len & Average length of comments in an issue \\
 & cm-developers-ratio & The ratio of number of comments to number of developers engaged in the discussion \\
 & time-to-discuss & The time span of the discussion \\\hline

\multirow{6}{*}{Events} & events & Number of events that happened to an issue\footnotemark \\
 & assigned & Is an issue assigned to a team member? \\
 & is-pull-request & Is an issue a Pull Request? \\
 & has-commit & Does an issue have any referenced commit? \\
 & has-milestone & Does an issue have a milestone? \\
 & labels   & Number of labels assigned to an issue \\\hline

\multirow{10}{*}{Developer} & author-followers & Number of followers of the author of an issue\\
 & author-following & Number of GitHub users the author follows \\
 & author-public-repos & Number of public repositories of the author of an issue \\
 & author-public-gists & Number of gists of the author of an issue \\
 & author-issue-counts & Number of issues opened by the author of an issue \\
 & author-github-cntrb & Number of contributions of the author of an issue in GitHub\\
 & author-account-age & The age of the author's GitHub profile account \\
 & author-repo-cntrb & Number of contributions of the author to the current repository\\
 & association & Association type of the author i.e., Collaborator, Contributor, Member, Owner, None \\
%  & closer-followers & Number of followers of the closer of an issue \\
%  & closer-following & Number of GitHub users the closer follows \\
%  & closer-public-repos & Number of public repositories of the closer of an issue \\
%  & closer-public-gists & Number of gists of the closer of an issue \\
%  & closer-github-cntrb & Number of contributions of the closer of an issue in GitHub \\
%  & closer-account-age & The age of the closer's GitHub profile account \\
%  & closer-repo-cntrb & Number of contributions of the closer of an issue to the current repository \\
 & same-author-closer & Are the author and closer same people? \\\hline

\multirow{4}{*}{Sentiment} & desc-positivity & Positive sentiment score of the description of an issue \\
 & desc-negativity & Negative sentiment score of the description of an issue \\
 & desc-pos-polarity & Positive polarity score of the description of an issue \\
 & desc-subjectivity & Subjectivity score of the description of an issue \\
\noalign{\smallskip}\hline
\end{tabular}
\end{table}
\footnotetext{A complete list of all issue events is available at \url{https://developer.github.com/v3/issues/issue-event-types/}}

\subsection{Feature Normalization}\label{sec:feat_norm}
As the value of our features selected in the previous step 
vary in degrees of magnitude and range, 
we perform feature normalization.
Machine Learning algorithms such as Logistic Regression (LR) and Neural Networks 
that use gradient descent as an optimization technique require data to be scaled.
Furthermore, distance-based algorithms like 
K-Nearest Neighbors (KNN), and Support Vector Machines (SVM) 
are affected by the range of features. 
This is because they use distances between data points 
to determine their similarity.
We use the \textit{Min-Max scaling} technique in which values of features 
are re-scaled to be in the fixed range of $0$ and $1$~\cite{al2006data}. 
We apply Min-Max scaling using Equation \ref{eq:minmax_scaling}
where $X_{max}$ and $X_{min}$ 
are the maximum and the minimum values of a feature, respectively.
We apply this technique to all our selected numerical features from the previous section.
\begin{equation}
\label{eq:minmax_scaling}
X_{norm} = \dfrac{X- X_{min}}{X_{max} - X_{min}}
\end{equation}

\subsection{Balancing Techniques}\label{sec:balancing}
A training dataset is considered to be imbalanced
if one or more of the classes are represented by
significantly less number of samples (issues) 
compared to other classes. 
This leads to skewed data distribution between classes 
and can introduce bias into the model \cite{weiss2001effect}.
Therefore, we employ two balancing techniques 
to improve the performance of our models.
We first assign higher weights to the less-represented classes.
The classifier is penalized based on these weights 
when it misclassifies issues.
The weight vector corresponding to our classes 
is calculated using Equation \ref{eq:class_weights}, 
where $ N$ is the number of issues in the whole dataset
and $frequency_{t_i}$ is the number of issues per class.
Second, we use the Synthetic Minority Oversampling Technique (SMOTE) 
\cite{chawla2002smote} to augment (over-sample) the minority classes.
% We use these techniques on the dataset of 
% the first task (predicting issue-objective labels) 
% as it is slightly unbalanced, 
% however, our second phase prioritizing) do not need balancing.
\begin{equation}
\label{eq:class_weights}
weight_{t_i} = \dfrac{N}{frequency_{t_i}}
\end{equation}

\subsection{Second Stage: Priority Prediction}\label{sec:model_priority}
We train our classifiers for the second stage of our approach.
To predict the priority level of issues,
we use our issue-priority dataset.
The input of the model in this phase 
consists of three different types of feature sets 
obtained from previous steps and explained below.
Table \ref{tab:model_inputs} summarizes the input to our model.
\begin{itemize}
    \item \textbf{Textual Features (TF)}: 
    First part of the input is the concatenated TF-IDF vectors of 
    title and description of an issue.
    We fit our TF-IDF vectorizer on the training dataset.
    Later we use the same vectorizer to transform the test dataset.
    We set the maximum number of features for 
    title and description vectors to $10K$ and $20K$.
    The objective label of an issue 
    which is the output of the first stage of the proposed approach 
    is also included.
Moreover, we set ngram range to $(1,2)$.
    \item \textbf{Labels Features (LF)}: 
    The second feature set is the one-hot vector of 
    available labels for an issue
    computed in Section \ref{sec:label_clusters}.
    \item \textbf{Normalized Features (NF)}: 
    And the third part of the input is the normalized version of 
    our engineered features obtained in Section \ref{sec:feat_norm}. 
    The complete list of selected features 
    is provided in Table \ref{tab:feature_selection}. 
    We also include sentiment scores in this set.
\end{itemize}
%%%%
\begin{table}[tb!]
\caption{Inputs to our models in both stages}
\label{tab:model_inputs}
\begin{tabular}{lp{85mm}}
\hline\noalign{\smallskip}
Stage & Inputs \\
\noalign{\smallskip}\hline\noalign{\smallskip}
Objective detection & 
- Word vectors of issue titles,

- Word vectors of issue descriptions. \\\midrule
Priority prediction & 
- TF-IDF vectors of issue titles,

- TF-IDF vectors of issue descriptions,

- Predicted objective of issues from the first stage, 

- One-hot encoded vector of available labels,

- Normalized feature vector 
    (containing five different set of information, namely textual information, discussion-related, developer-related, events-related, and sentiment scores).
\\
\noalign{\smallskip}\hline
\end{tabular}
\end{table}

We use RF as the selected classifier in this phase. 
RF has been shown to perform better on tabular data.
We configure the model parameters
using \textit{Random Search} algorithm~\cite{bergstra2012random}, 
that performs an exhaustive search of
the hyper-parameter space.
The output of the second stage is the probability of an issue to be High or Low.
Note that we use the two balancing techniques 
presented in Section~\ref{sec:balancing}  
to improve the performance of our classifier, 
for the project-based setting, 
where for some repositories, 
the ratio of labels is very unbalanced.

%%%%%%%%%%%%%%%%%%%%%%% Evaluation Setup %%%%%%%%%%%%%%%%%%%%%%%%%%%%%%%%
%%%%%%%%%%%%%%%%%%%%%%%%%%%%%%%%%%%%%%%%%%%%%%%%%%%%%%%%%%%%%%%%%%%%%%
\section{Experiment Design}\label{sec:expriment}
We conduct various experiments to validate the proposed approach.
An experiment is designed to analyze the performance of the issue-objective predictor model.
For the priority prediction task, 
we analyze the model in both project-based and cross-project settings.
Our priority prediction model has two applications: 
(1) to prioritize open or closed issues 
to facilitate timely task assignment and better project management, 
(2) to help select important issues for inclusion in the periodic documentation of the project, 
e.g., to automatically select important changes for inclusion in release notes.
It is worth mentioning that all issue features
in both experiments of the project-based and cross-project settings 
for the priority prediction task
are calculated after issues are closed.
However, a possible future research direction 
is to design an experiment for training and evaluating
the priority prediction model  
through collecting dynamic features periodically.
We also conduct a human labeling and evaluation experiment on unlabeled issues.

We use the datasets presented in Section~\ref{sec:data_collection}.
We split them to train, and test sets 
with ratios of $80\%$, and $20\%$.
Note that we use stratified sampling on the target value to randomly select these datasets
to reduce sampling biases and retain the similar class distribution in train, test and the whole dataset.
On smaller datasets, we also incorporate cross-validation technique.

We conduct our experiments on a machine with Ubuntu 16.04, 64-bit,
Intel(R) Xeon(R) CPU E5-2690 3.00GHz and 64.0GB RAM.
Next, we present our Research Questions (RQ) 
and the performance metrics for 
evaluating our model on the collected datasets.

\subsection{Research Questions}
In this study, we investigate the following research questions:
\begin{itemize}
    \item \textbf{RQ1}: How accurately our model predicts the objective behind opening an issue?
    
    We train a text classifier on 
    a large-scale dataset of $818$K issue reports 
    to investigate how accurately we can predict the objective of an issue. 
    The list of objectives that we consider in this phase 
    are among the most-used labels in GitHub, namely Bug, Enhancement, and Support.
    \item \textbf{RQ2}: How accurate is our priority prediction model in a project-based context?
    
    We train classifiers for each repository separately.
    The goal is to predict the importance of an issue.
    This predicted label can then be used 
    for prioritizing team resources 
    for solving the high-priority issues 
    or used for documentation purposes of the project.
    One use case of the latter are release notes 
    (or any other types of reports).
    That is, team managers, tasked with writing reports of each release, 
    can use the model to extract the urgent issue reports 
    addressed by the team for that release.
    \item \textbf{RQ3}: How accurate is our priority prediction model in a cross-project setting?
    
    That is how well does our trained classifier performance 
    transfer to other repositories? 
    We investigate the previous research question but in a cross-project setting. 
    We train our model on $80\%$ of repositories and investigate how well this generic model 
    predicts the priority label of issues from other repositories?
    
    \item \textbf{RQ4}: How does the priority prediction model preform on unlabeled data?
    
    We conduct human labeling and evaluation 
    to assess the performance of the priority detection model trained 
    in a cross-project context on unlabeled issues from unseen GitHub projects. 
    Moreover, through an open question, we ask 
    what are the factors participants take into account when categorizing issues into high and low priority.
\end{itemize}

\subsection{Evaluation Metrics}
We use standard measures for evaluating classifiers, 
namely \textit{Precision}, \textit{Recall}, \textit{F1-score} and Accuracy.
Precision computes the percentage of issues that are
correctly predicted with label X 
over all the issues classified as X.
Recall, on the other hand, computes
the percentage of issues that are
correctly predicted as X over all the issues labeled as X in our ground-truth.
Fl-score is the harmonic mean of these two.
Finally, we also report \textit{Accuracy} which is
the ratio of correct predictions, 
both true positives (TP) and true negatives (TN), 
from the total number of cases examined.
TP indicates the number of truly X-labeled issues that are classified as X.
FP is the number of truly Non-X issues that are classified as X.
True Negative (TN) denotes the number of truly Non-X issues that are classified as Non-X.
And False Negative (FN) indicates the number of truly X-labeled issues that are classified as Non-X.
Equations \ref{eq:precision}, \ref{eq:recall},
\ref{eq:f1}, and \ref{eq:accuracy} compute the above measures.
\begin{equation}
\label{eq:precision}
Precision = \frac{\text{TP}}{\text{TP + FP} }
\end{equation}
\begin{equation}
\label{eq:recall}
Recall = \frac{\text{TP}}{\text{TP + FN} }
\end{equation}
\begin{equation}
\label{eq:f1}
F1 = \frac{2 \cdot Precision\cdot Recall}{Precision+ Recall}
\end{equation}
\begin{equation}
\label{eq:accuracy}
Accuracy = \frac{TP+TN}{TP+FP+TN+FN}
\end{equation}

\subsection{Baselines}
For both tasks, 
we include baselines from a wide range of rule-based and learning-based solutions.

For the first task, objective detection, 
we train several supervised machine learning-based models
on a large-scale dataset of preprocessed $818$K issue reports 
to predict their objective (Bug Report, Enhancement, or Support/Documentation).
We use TicketTagger~\mbox{\cite{kallis2019ticket,song2020bee}}
and Intention-mining~\mbox{\cite{huang2018automating}}
as the baselines for this task.
Moreover, we train two more classifiers based on Multinomial Naive Bayes  
and Bidirectional Long Short Term Memory (BiLSTM) deep neural networks
that are usually used for text classification
as complementary baselines.
The latter is inspired by the study of Li et al.~\mbox{\cite{li2019tagdeeprec}} 
for tag recommendation in software information sites.
Finally, we also implement a keyword-based approach 
to include simpler rule-based solutions in the experiment.
In this baseline, the model looks for specific keywords 
related to the three categories above 
and tags them with their respective label.
For example, if the issue contains words 
such as \textit{crash} and \textit{fix}, 
it will be labeled as a bug report.

For the second task, priority prediction,
we include baselines 
which are all vanilla (standard) versions of classical Machine Learning models, 
namely KNN, Multinomial NB, Logistic Regression and RF.
Furthermore, we add several simpler models 
based on the date of issues or number of comments to the list of the baselines for this task.
For instance, for the ``Most Comments" baseline, 
we calculate the median number of comments for issues. 
We then proceed to tag those with a higher number of comments 
than the median value with HP and the rest with LP labels.
Finally, we also include the proposed approach 
by Dhasade et al.~\mbox{\cite{dhasade2020towards}}, 
\textit{Issue Prioritizer}, for this task.

\subsection{Human Labeling and Evaluation: Setup}
We designed an experiment to investigate 
whether the trained model in the cross-context setting 
can be used successfully 
for labeling \textit{unlabeled} issue reports.
As this experiment is designed 
to assess the performance of the proposed model on unlabeled issues,
we do not have the ground truth labels to compare against.
Thus, we employ a \textit{partially objective labeling task}~\mbox{\cite{alonso2014crowdsourcing}}, 
a crowd-sourced labeling task in which 
the label (High or Low Priority) of a subject (issue report)
is determined based on inter-rater agreement among the participants.
That is, a given issue report is assigned the label  
which the majority of raters have given it.
We then compare these majority-vote labels with the labels generated 
by our priority detection model.

We provided general information about each project for the participants 
to help them make informed decisions. 
This information include 
the project's goal, description, \#stars, \#forks, \#contributors, \#closed and \#open issues, 
and median response time by the developers of the project to its issues.
Furthermore, participants were instructed to analyze the assigned repository 
and its main characteristic to get familiar with the project.
We asked the participants to assess ten issues of a given project 
and assign a High or Low priority label 
to each one based on the characteristics of the project.
Next, with an open question, 
we asked what factors participants took into account 
while tagging the issues.

\paragraph{Projects} 
We randomly selected $60$ issue reports from 
six unseen GitHub projects (ten issues per project).
The list of projects is as following. 
They were selected based on their popularity, and the variety in their sizes.
Moreover, all projects' main programming language is Java.
\begin{itemize}
    \item \textbf{Elasticsearch}: Free and Open, Distributed, RESTful Search Engine,
    \item \textbf{Spring Boot}: Spring Boot makes it easy to create stand-alone, production-grade Spring based Applications that you can just run,
    \item \textbf{OkHttp}: Square's meticulous HTTP client for the JVM, Android, and GraalVM,
    \item \textbf{RxJava}: Reactive Extensions for the JVM; 
    a library for composing asynchronous and event-based programs using observable sequences for the Java VM,
    \item \textbf{Retrofit}: A type-safe HTTP client for Android and the JVM,
    \item \textbf{Guava}: Google core libraries for Java.
\end{itemize}

\paragraph{Participants}
As we did not have access to the main developers of these projects, 
we invited $62$ Software Engineers 
from both industry and academia to participate in this study.
Each participant was assigned to the issues of one project.
Thirty four Software Engineers responded and participated in the study
from which $30$ responses were valid ($25$ males and $5$ females).
Thus our response rate is $48\%$.
All participants have a BSc or MSc in Software Engineering 
with average of $4.8$ years of developing experience.
They all are proficient with the programming language Java.
In addition, on average the participants own 
or contribute to $6.52$ open-source projects on GitHub.

\paragraph{Inter-rater Reliability Measurement}
Inter-rater reliability
is the level of agreement among independent observers 
who label, code, or rate the same phenomenon~\mbox{\cite{gwet2008computing}}.
Several statistics can be used to measure inter-rater reliability, 
from which the most common are \textit{Percent Agreement}, 
\textit{Cohen's kappa} (for two raters), 
and \textit{Fleiss kappa} as an adaptation of Cohen's kappa for three or more raters~\mbox{\cite{fleiss1973equivalence}}.
To compute Percent Agreement score among the participants, 
we create a matrix in which the columns represented the different labelers, 
and the rows represent issue reports. 
The cells of this matrix contain the label (category) the labelers entered for each issue.
As we only have two labels (High and Low Priority), 
we fill the cells with either $0$ (Low Priority) or $1$ (High Priority)
For each row, we calculate the the Percent Agreement and then report the average.
Percent agreement ranges between $0$ and $1$, 
with $0$ as no agreement and $1$ as perfect agreement.
% Fleiss' kappa is a generalisation of Scott's pi statistic~\mbox{\cite{fleiss2013statistical}},
% and is used to assess the reliability of agreement between 
% any number of raters (more than two) when classifying items. 
Kappa determines the extent to which the observed amount of agreement among labelers 
surpass the expected value if all labelers tagged issues completely randomly.
Two variations of kappa for multi-raters (more than two)
are Fleiss' fixed-marginal multi-rater kappa~\mbox{\cite{fleiss1971measuring}} 
and Randolph's free-marginal multi-rater kappa~\mbox{\cite{randolph2005free}}.
Marginal distributions are considered to be \textit{free} 
when the quantities of cases that should be distributed into each category is not predefined.
As our labelers are not forced to assign a fixed number of issues to each label (category)
we report Randolph's free-marginal kappa score for this experiment~\mbox{\cite{brennan1981coefficient}} 
Values of kappa can range from $-1.0$ (perfect disagreement  below chance),
to $0$ (agreement equal to chance), 
to $1.0$ (perfect agreement above chance). 
The kappa will be higher when there are fewer categories.
% Fleiss et al.~\mbox{\cite{fleiss2013statistical}}  
% suggest the following score system for interpreting kappa values
% less than $0.4$ as poor agreement,
% $0.4 – 0.75$ as	good agreement, and
% more than $0.75$ as almost perfect agreement.
Landis and Koch.~\mbox{\cite{landis1977application}}  
suggest the following system for interpreting kappa values:
\begin{itemize}
    \item less than $0$ as poor agreement,
    \item $0.01 – 0.20$ as slight agreement, 
    \item $0.21 – 0.40$ as fair agreement, 
    \item $0.41 – 0.60$ as moderate agreement, 
    \item $0.61 – 0.80$ as substantial agreement, and
    \item $0.81 – 1.0$ as almost perfect agreement. 
\end{itemize}
%%%%%%%%%%%%%%%%%%%%%%% Results %%%%%%%%%%%%%%%%%%%%%%%%%%%%%%%%%%%%%%%%%
%%%%%%%%%%%%%%%%%%%%%%%%%%%%%%%%%%%%%%%%%%%%%%%%%%%%%%%%%%%%%%%%%%%%%%

\section{Experiment Results}\label{sec:results}
In the following, we report the results of our experiments 
and the answer to our research questions.

\subsection{RQ1: Issue Objective Detection}
Table \ref{tab:rq1-results} reports the results of objective-prediction task.
$B$, $E$, and $SD$ represent Bug Report and Enhancement and Support/Documentation classes.
As presented, our proposed approach indeed has a high accuracy for predicting issue-objective labels.
We successfully outperform all the baselines regarding all the evaluation metrics.
For instance, regarding F1-score of the Support class,
we outperform
BiLSTM, 
Multinomial NB, 
CNN (Intention mining), 
and FastText (TicketTagger and BEE) -based classifiers
by $204\%$, $66\%$, $43\%$, $63\%$, and $24\%$, respectively.
The keyword-based approach do not achieve sufficient accuracy. 
That is probably because an issue can contain prominent but conflicting keywords. 
For instance, a user can describe a bug, but does not use the bug-related vocabulary explicitly, 
hence misleading such simple models.
Furthermore, although Hunag et al.~\mbox{\cite{huang2018automating}} 
use a deep model for classification, 
it is not performing very well in our case as is.
The problem can be due to the fact that their model 
is designed and optimized to predict the goal of each sentence of an issue report separately.
As we have not adapted the architecture of their approach to our goal, 
it may not be suitable for predicting the objective of a complete issue report.
However, it can provide a preliminary analysis 
on using different Convolutional deep neural networks in classifying issues.
Finally, the BiLSTM-based deep model also does not perform well in our case 
and it takes a very long time to train as expected 
when training Recurrent Neural Networks on large datasets.
On the contrary, the inherent parallelization of Transformers' architecture
allows our proposed approach to be trained much faster along providing better results.

While we outperform all the baselines for all classes, 
our model too seems to struggle for the \textit{Support/Documentation} class
(compared to the other two classes; Bug report, and Enhancement).
This is probably due to several reasons including 
(1) this class is under-presented in our dataset (less number of issues), 
and (2) the objective behind opening issues in this category is inherently more diverse.
As described in Section \ref{sec:data_collection}, 
we include various issues tagged by labels 
such as \textit{question}, \textit{support}, \textit{help}, etc. 
in the Support/Documentation category. 
While Kallis et al.~\cite{kallis2019ticket} only consider 
issues tagged with the label \textit{question} as their third class,
our goal is to cover a broader range of issue reports in the third class 
and provide a more generic objective classifier.
It is also worth mentioning, 
usually fixing existing bugs or implementing requested features 
are of more value to the community, 
thus, we believe better performance on the first two objectives (bugs and enhancement) is deemed more important.
Nonetheless, one can improve these models' performance by collecting more data for this class 
or narrowing down the objective of this class.
It is worth mentioning that 
we investigated the use of various additional features for this task. 
As the results did not improve significantly,
we decided to keep it simple for the first stage 
and only incorporate textual information of issues.
\begin{table}[tb!]
\caption{RQ1: Objective detection results}
\label{tab:rq1-results}
\begin{tabular}{lllll|lll|lll}
\hline\noalign{\smallskip}
& \multicolumn{10}{c}{Evaluation metrics}\\
\cmidrule{2-11}
  & Accuracy 
& \multicolumn{3}{c}{Precision} 
& \multicolumn{3}{c}{Recall} 
& \multicolumn{3}{c}{F1-score} \\
\cmidrule{2-11}
Model / Classes &  
& B & E & SD 
& B & E & SD  
& B & E & SD \\
\noalign{\smallskip}\hline\noalign{\smallskip}
Keyword-based & 26\%
& 66\% & 63\% & 26\% & 39\% & 15\% & 19\% & 49\% & 25\% & 22\%\\
Multinomial NB & 73\% 
& 79\% & 71\% & 62\% & 75\% & 83\% & 37\% & 77\% & 77\% & 47\%\\%un-tune
Bi-LSTM & 68\% %accuracy
& 71\% & 71\% & 48\% %precision
& 77\% & 72\% & 34\% %recall
& 74\% & 72\% & 40\%\\ %f1
CNN~\cite{huang2018automating} & 73\% 
& 74\% & 73\% & 54\% & 80\% & 77\% & 32\% & 77\% & 75\% & 41\%\\
FastText~\cite{kallis2019ticket,song2020bee} & 76\% 
& 78\% & 77\% & 67\% & 82\% & 80\% & 46\% & 80\% & 78\% & 54\%\\
\midrule
Proposed approach & \textbf{82\%} 
& \textbf{84\%} & \textbf{83\%} & \textbf{72\%} 
& \textbf{86\%} & \textbf{84\%} & \textbf{62\%} 
& \textbf{85\%} & \textbf{84\%} & \textbf{67\%}\\
\noalign{\smallskip}\hline
\end{tabular}
\end{table}

\subsection{RQ2: Project-based Priority Prediction}
In this experiment, 
we train our proposed approach per repository to predict the priority of their issues.
As discussed in Section~\ref{sec:model_priority}, 
we have three sets of input features, namely TF, LF, and NF 
and experimented with different combinations of them.
For each repository, we take $80\%$ of its issues as the train data 
and test the model on the remaining $20\%$.

Table~\ref{tab:rq2-results} reports 
the results of the project-based priority-prediction task 
on these $70$ repositories. 
As mentioned before, $HP$ and $LP$ represent High-Priority and Low-Priority classes.
As there are various feature sets attributed to an issue report, 
we investigated the use of different variations of these feature sets 
and report the best case below.
The results indicate integrating selected features (refer to Section~\ref{sec:model_priority})
helps training a better model.
As shown, our proposed approach based on RF
with all the three input feature vectors (NF, LF and TF) 
outperforms all the baselines.
This indicates the benefit of integrating other features, 
and employing normalization, balancing, and optimization techniques.
Lastly, our experiments also show 
both of the balancing techniques proposed in Section~\ref{sec:balancing}
perform comparably.
\begin{table}[tb!]
\caption{RQ2: Project-based priority prediction results}
\label{tab:rq2-results}
\begin{tabular}{llll|ll|ll}
\hline\noalign{\smallskip}
& \multicolumn{7}{c}{Evaluation metrics}\\
\cmidrule{2-8}
& Accuracy 
& \multicolumn{2}{c}{Precision} 
& \multicolumn{2}{c}{Recall} 
& \multicolumn{2}{c}{F1-score} \\
\cmidrule{2-8}
Models / Classes &  
& HP & LP
& HP & LP
& HP & LP\\
\noalign{\smallskip}\hline\noalign{\smallskip}
Oldest & 48\% 
& 43\% & 52\% 
& 48\% & 47\%  
& 44\% & 49\%\\
Recently updated & 48\% 
& 44\% & 53\% 
& 48\% & 50\%  
& 45\% & 50\%\\ 
Most comments & 50\% 
& 46\% & 54\% 
& 64\% & 39\%  
& 52\% & 44\%\\\midrule
Issue Prioritizer~\cite{dhasade2020towards} & 55\% 
& 58\% & 53\% 
& 51\% & 61\%  
& 54\% & 57\%\\
KNN & 67\% 
& 63\% & 69\% 
& 64\% & 70\%  
& 63\% & 68\%\\
Multinomial NB & 69\% 
& 68\% & 69\% 
& 63\% & 77\%  
& 60\% & 69\%\\
Logistic Regression & 70\% 
& 69\% & 71\% 
& 66\% & 76\%  
& 64\% & 71\%\\
Vanilla RF & 69\% 
& 69\% & 69\% 
& 65\% & 75\%  
& 63\% & 70\%\\\midrule
Proposed approach & \textbf{75\%} 
& \textbf{73\%} & \textbf{77\%}
& \textbf{74\%} & \textbf{78\%}  
& \textbf{72\%} & \textbf{77\%} \\
\noalign{\smallskip}\hline
\end{tabular}
\end{table}

\begin{figure}[tb!]
    \centering
    \includegraphics[width=\textwidth]{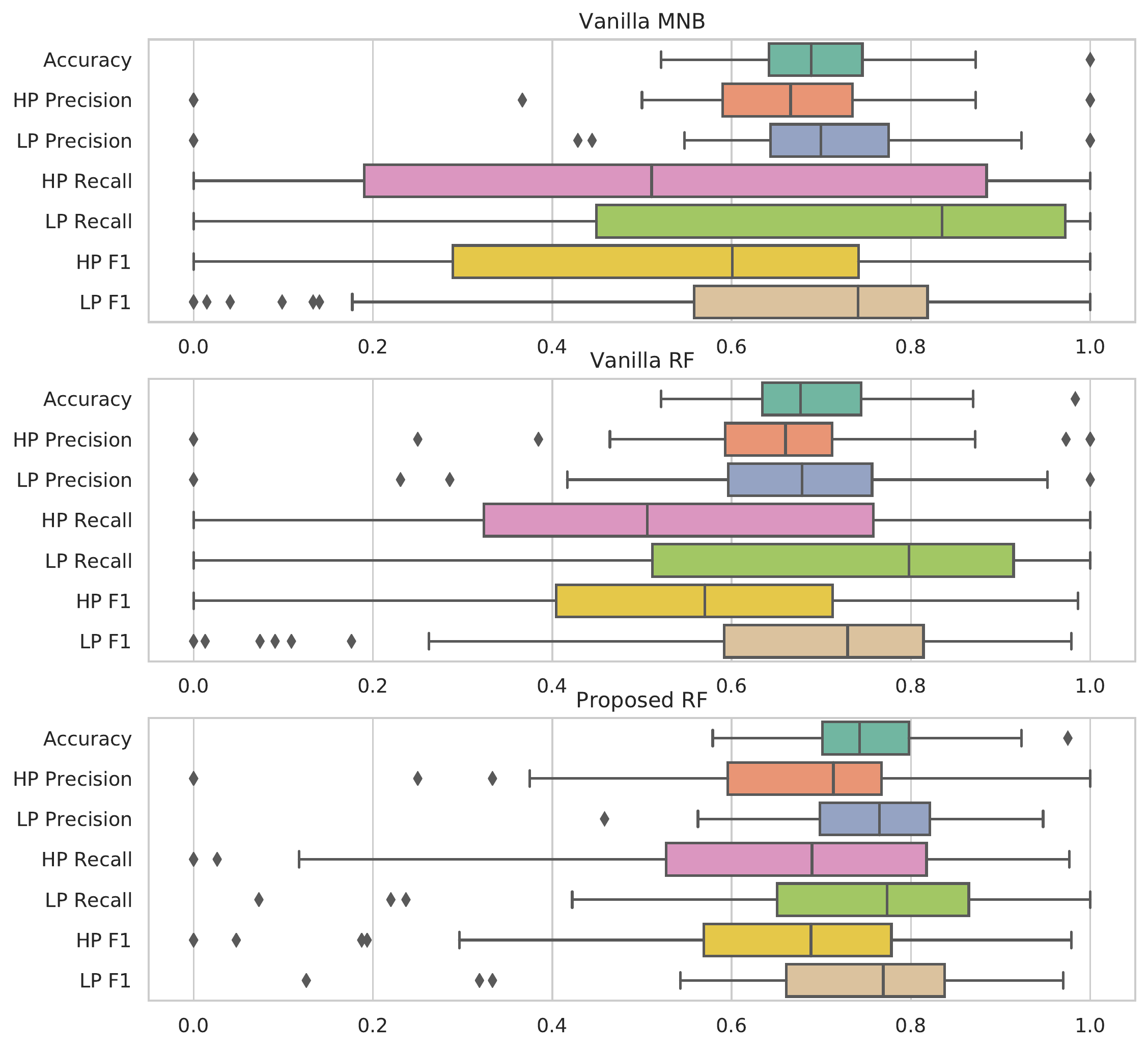}
    \caption{Distribution of results among $70$ repositories for three approaches.}
    \label{fig:rq2_boxplots}
\end{figure}

Figure \ref{fig:rq2_boxplots} provides three sets of box-plots for three approaches 
including two baselines and one of our proposed approach based on RF.
The box-plots report the distribution of results per repository 
and based on all the evaluation metrics.
Comparing vanilla Multinomial NB and our approach, 
it is clear that enriching these classifiers 
with the advance techniques mentioned in Section \ref{sec:model_priority}
cause the model to perform more consistently.
For instance, take the box-plot 
for the Recall metric (for the HP class) 
provided by Multinomial NB and our approach.
Our approach scores above $50\%$ for all repositories, 
while the Multinomial NB performance fluctuates for different repositories 
with the first quartile as low as $20\%$.
In fact, the first quartile for all the metrics 
and all the repositories are above $50\%$ in our approach.
Moreover, comparing the vanilla RF with our enriched version of RF, 
one can see that the latter shifts the results of all metrics to higher scores.
Therefore, we are able to successfully 
outperform the baselines regarding different evaluation metrics.

\subsection{RQ3: Priority Prediction in Cross-project Context}
We also train a generic model on issues from $80\%$ of repositories 
and evaluate this model on the rest of the repositories.
Our goal is to investigate whether a generic model trained in a cross-project setting
can perform on par with project-based models.
Table~\ref{tab:rq3-results} reports the results of priority-prediction task 
in the cross-project context. 
Based on the results, our proposed approach based on RF 
with the two feature inputs (NF, and LF) outperforms all other models. 
% More specifically, we outperform the state of the art, Issue Prioritizer, by 35\%.
This generic model can indeed perform comparably 
with the average performance of project-based models.
Therefore, we can train only one generic model 
to automatically predict the priority of issues
and successfully reuse (and/or retrain) it for unseen repositories 
or repositories with little historical data.
It is worth mentioning that 
in our case, TF-IDF vectors provide more information to these models 
compared to Doc2Vec and Word2Vec vectors.
\begin{table}[tb!]
\caption{RQ3: Cross-project priority prediction results}
\label{tab:rq3-results}
\begin{tabular}{llll|ll|ll}
\hline\noalign{\smallskip}
& \multicolumn{7}{c}{Evaluation metrics}\\
\cmidrule{2-8}
& Accuracy 
& \multicolumn{2}{c}{Precision} 
& \multicolumn{2}{c}{Recall} 
& \multicolumn{2}{c}{F1-score} \\
\cmidrule{2-8}
Models / Classes &  
& HP & LP
& HP & LP
& HP & LP\\
\noalign{\smallskip}\hline\noalign{\smallskip}
Oldest & 47\% 
& 43\% & 51\% 
& 47\% & 47\% 
& 45\% & 49\%\\
Recently updated & 50\% 
& 46\% & 54\% 
& 50\% & 50\% 
& 48\% & 52\%\\
Most comments & 50\% 
& 47\% & 55\% 
& 53\% & 49\% 
& 50\% & 52\%\\
\midrule
KNN & 57\% 
& 47\% & 63\% 
& 38\% & 71\% 
& 42\% & 67\%\\
SVM & 58\% 
& 47\% & 64\% 
& 43\% & 68\%  
& 45\% & 66\%\\
Logistic Regression & 57\% 
& 46\% & 64\% 
& 45\% & 65\% 
& 46\% & 64\%\\
Multinomial NB & 62\% 
& 55\% & 66\% 
& 40\% & 78\% 
& 46\% & 71\%\\
Issue Prioritizer~\cite{dhasade2020towards} & 55\% 
& 55\% & 55\% 
& 50\% & 60\% 
& 53\% & 57\%\\
Vanilla  RF & 57\% 
& 46\% & 63\% 
& 43\% & 67\% 
& 44\% & 65\%\\
\midrule
Proposed approach & \textbf{74\%} 
& \textbf{70\%} & \textbf{75\%} 
& \textbf{59\%} & \textbf{83\%}  
& \textbf{64\%} & \textbf{79\%} \\
\noalign{\smallskip}\hline
\end{tabular}
\end{table}

\subsubsection{Feature Importance}
Using RF, we derived the importance of 28 features listed in \autoref{tab:feature_selection} (NF feature vector).
The five most important features are 
\textit{time\_to\_discuss}, 
\textit{cm\_mean\_len}, \textit{desc\_words}, \textit{desc\_subjectivity}, and \textit{desc\_pos\_polarity}.
The least important five features are 
\textit{numeric\_association}, \textit{code}, \textit{has\_commit}, \textit{same\_author\_closer}, and finally \textit{is\_pull\_request}.
Moreover, the five most important features from the LF feature vector are
\textit{bug}, \textit{feature}, \textit{documentation}, \textit{stale}, \textit{usability}, \textit{won't fix}. 
The least five features from this vector are \textit{weekly-report}, \textit{announcement}, \textit{pinned}, \textit{hard}, \textit{bounty}.

\subsubsection{High vs. Low priority}
\begin{figure}[tb!]
    \centering
    \includegraphics[width=0.65\textwidth]{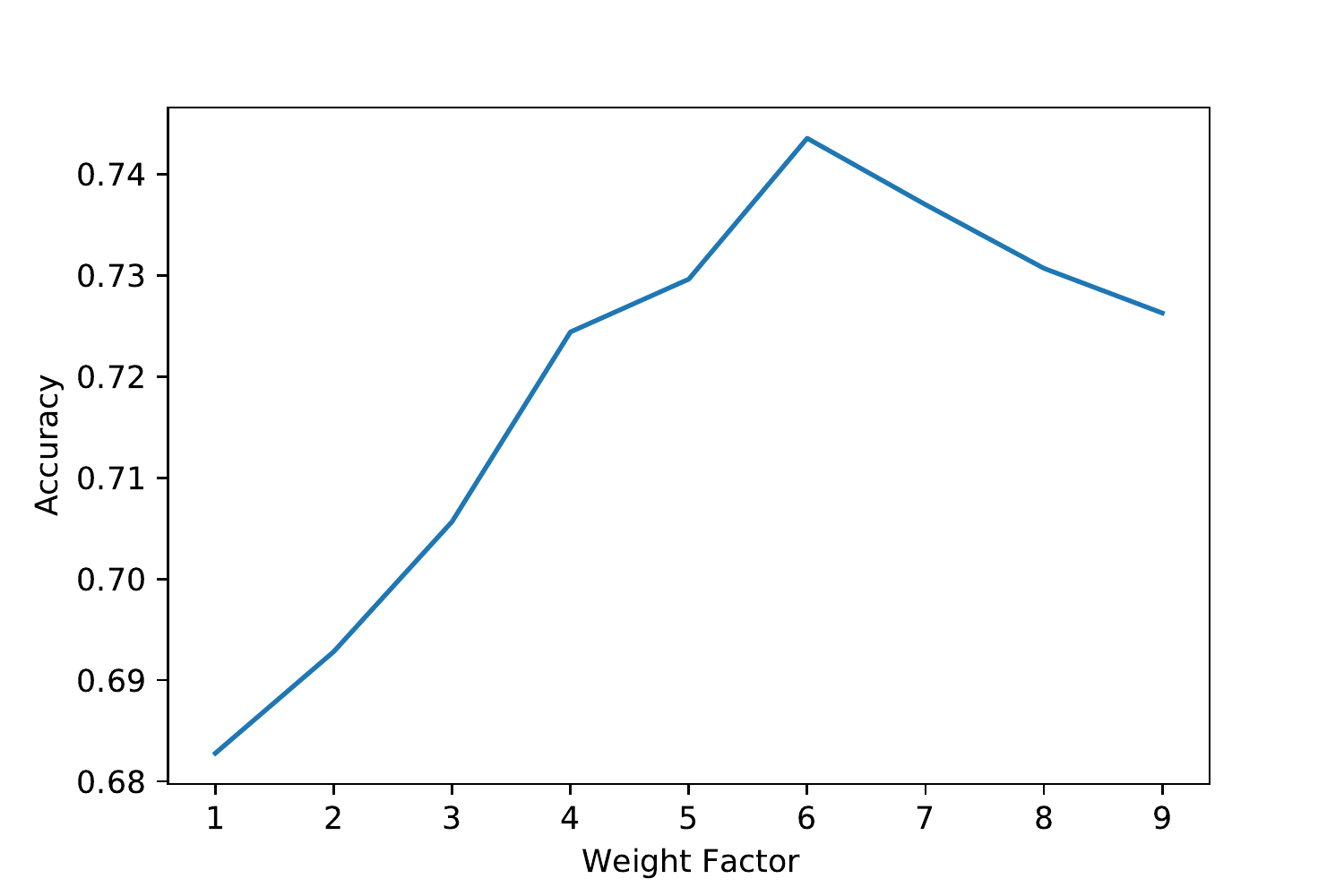}
    \caption{The impact of class weights on total accuracy}
    \label{fig:class_w_acc}
\end{figure}

\begin{figure}[tb!]
    \centering
    \subfigure[HP class]
    {\includegraphics[width=0.49\linewidth]{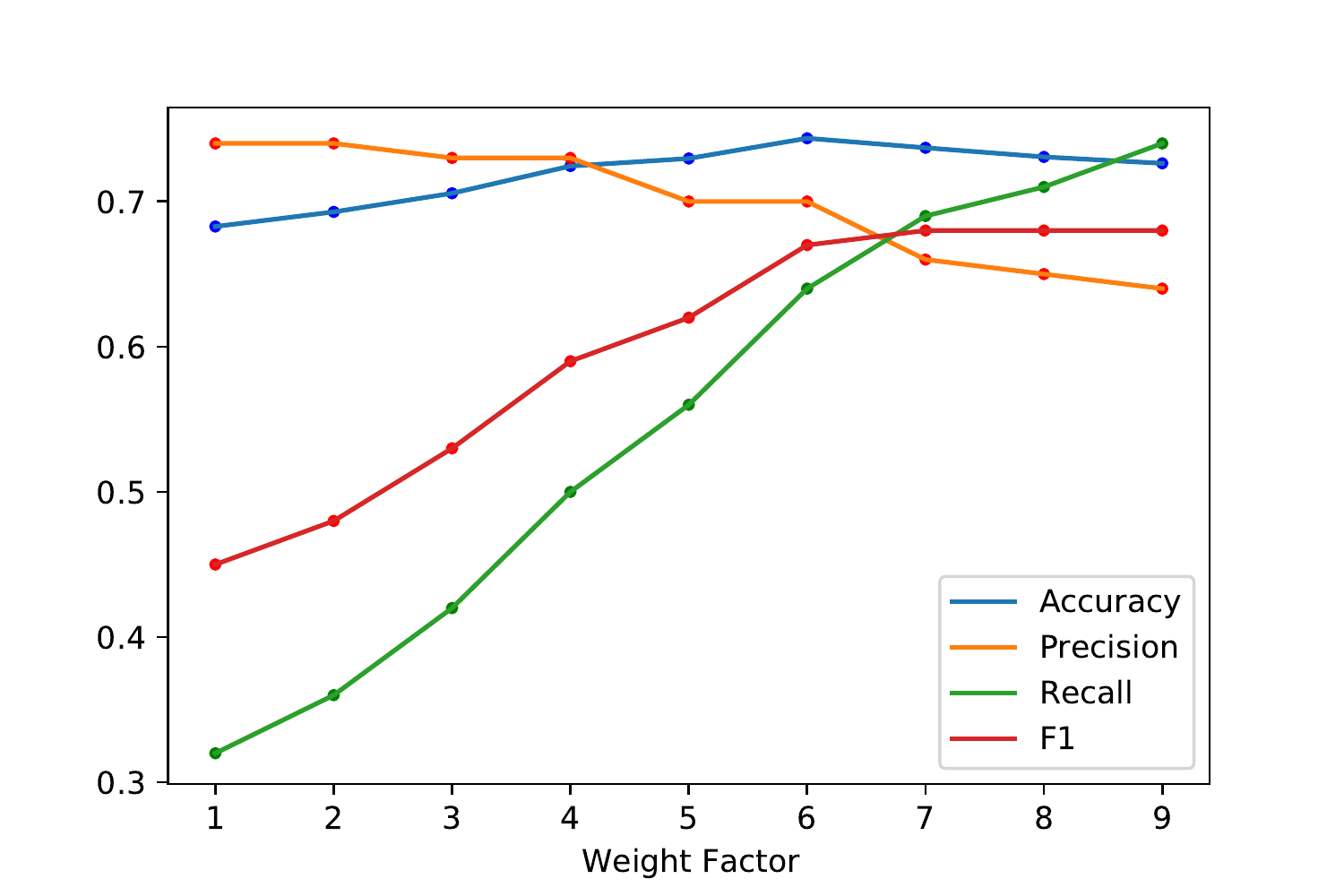}}
        %\label{fig:class_w_hp}
    \subfigure[LP class]
    {\includegraphics[width=0.49\linewidth]{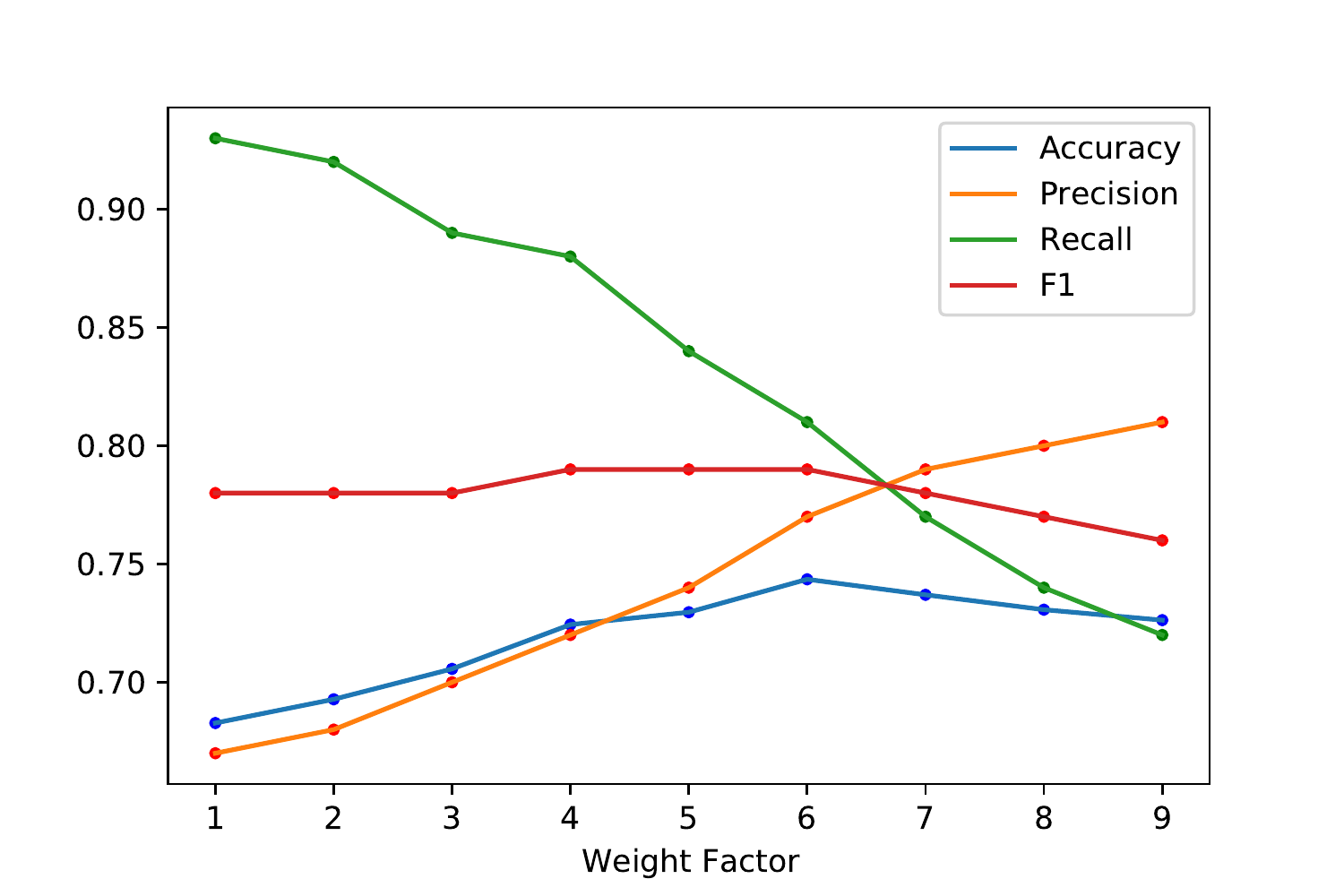}}
        %\label{fig:class_w_lp}
    \caption{The impact of changing class weights per class}
    \label{fig:class_w_LP_HP}
\end{figure}

As the impact of misclassifying HP and LP classes differ, 
they can be treated differently. 
In this section, we investigated the impact of adding more weights 
to the HP class to see how it affects the results.
To this end, we set the weights to 
$0.1 \times (10-i)$ for the LP class, 
and $0.1 \times i$ for the HP class where $i = [1, 9]$.
Figures \ref{fig:class_w_acc}, and \ref{fig:class_w_LP_HP} 
depict changes in the total accuracy, 
and results of different evaluation metrics for HP and LP classes.
The results indicate that 
as the weight of the HP class increases, the scores of recall, F1, and accuracy for HP class and precision for LP class increase. 
At the same time, precision of HP class, and Recall of LP class decrease.
However, the best overall accuracy of the model based on both classes 
is achieved when $i = 6$. 
That means slightly more emphasis on the HP class results in the best overall result.
As we have tuned the parameter, this is the setting we have used for the cross-project context as well.
That is, the final model puts more emphasis on the HP class to achieve the best overall results. 
However, in cases where the HP class (or LP) is of much higher importance, 
one can adjust the weights to get the desired results from the model.
%
% \begin{figure}[tb!]
%     \centering
%     \includegraphics[width=0.75\textwidth]{images/classwieght_HP_tuned.pdf}
%     \caption{The impact of class weights on the results of HP class}
%     \label{fig:class_w_hp}
% \end{figure}
%
% \begin{figure}[tb!]
%     \centering
%     \includegraphics[width=0.75\textwidth]{images/classwieght_LP_tuned.pdf}
%     \caption{The impact of class weights on the results of LP class}
%     \label{fig:class_w_lp}
% \end{figure}

\subsection{RQ4: Human Labeling and Evaluation: Results}
For the first part of this experiment, 
we asked participants to tag unlabeled issues from six unseen projects.
We collected at least $5$ votes (priority labels) per issue ($300$ votes in total).

We initially obtained $34$ responses to our questionnaires, 
from which four responses had major conflicts with others (outliers). 
We define an outlier labeler as an individual whose tagged labels 
are different than labels assigned by the majority of other labelers 
(who tagged the same issues) in more than $50\%$ of cases. 
To avoid introducing noise, we removed all the labels assigned 
by such outlier participants and then proceeded 
to assess the results based on the responses of 
the remaining $30$ individuals.  
The average outlier percentage for the remaining $30$ labelers, 
is $19\%$. 
That is, on average, a labeler, 
has assigned similar labels to what others tagged for the same issues 
in more than $80\%$ of cases.

The model achieves $90\%$ overall accuracy.
Moreover, accuracy per project ranges from $80\%$ to $100\%$.
Weighted precision, recall, and F1 scores for the two classes of HP and LP are
$92\%$, $90\%$ and $91\%$, respectively.
The results indicate the model is capable of predicting unseen issues successfully.
Note that the above accuracy is achieved using the cross-project-based priority prediction model.
We believe adding historical data of projects in GitHub 
and training project-based models can further improve these results.

To measure the inter-rater reliability among the $30$ participants, 
we use two measures, 
\textit{Percent Agreement} and \textit{Randolph's free-marginal multi-rater kappa}. 
We achieve $85.3\%$ overall Percent Agreement and $0.71$ Kappa. 
For the latter, based on Landis and Koch's interpretation system.~\mbox{\cite{landis1977application}},
the achieved score 
translates to \textit{substantial agreement} among the labelers. 
Thus, our labelers substantially agree according to this measure.
We also compute these measures per project. 
Percent Agreement among the six projects varies between $76\%$ to $96\%$.
Moreover, for two out of six projects (ElasticSearch and Retrofit), 
kappa is above $0.81$ 
which translates to \textit{almost perfect agreement} 
among labelers for these projects. 
Considering the diversity among participants, 
the large number of labelers, 
and the inherent subjectiveness when prioritizing issues, 
we believe the assigned labels have good quality and the labelers 
are mostly in agreement with each other. 
However, the exact same results may not be replicated using another set of labelers or issues.

In the second part of this experiment, 
we asked the participants what factors they take into account when determining the importance of issue reports.
In the following, we have summarized their free-format answers in several major groups to provide insights for future work.
Two of the authors were involved in the process of analyzing the free-format answers. 
We used the \textit{open coding} technique for this process by breaking issues into discrete parts and creating codes to label them~\mbox{\cite{khandkar2009open}}. 
Each author separately labeled each sentence of a free-format answer for all participants. 
We used concise summarization of a sentence's goal in the labeling process. 
If sentences were compound, authors separately labeled the goal of each phrase. 
Then, for each participant, the two authors compared the goal category for each sentence/phrase. 
In the end, we accumulated the categories, clustered them, 
and reported on the largest clusters existing in the data of this experiment.
While we have exploited some of these factors (e.g., issue type, discussion magnitude, roles, etc.) 
in this study,
other interesting factors such as 
the required effort and estimated impact 
can be also utilized to further improve the proposed model.
\begin{itemize}
    \item \textbf{Issue type}: 
    Many participants indicated that they first look for the type of issue, 
    whether it is a bug report, a feature request, a question, etc. 
    Then they go deeper, if it is a bug, what kind of a bug it is, e.g., is it security related? 
    \item \textbf{Content}: 
    Is it related to the core features of the project?
    \item \textbf{Impact}: 
    Is the reported issue blocking other functionalities of the project? 
    Is it affecting many users?
    Which one can potentially cause more problems?
    \item \textbf{Discussion/Reaction magnitude}: 
    How large is the discussion around the reported issue?
    How many comments has it attracted? 
    How many users are participating in the discussion? 
    What are the emojis used? 
    \item \textbf{Labels}: 
    What are the labels assigned to the issue, e.g., duplicate, invalid, etc? 
    \item \textbf{Roles}: 
    Who has opened the issue?
    What is their experience level?
    Which team members are participating in the discussion?
    \item \textbf{Required effort}: 
    How much effort is required to solve the issue?
    \item \textbf{Dates}: 
    How long ago has it been reported?
\end{itemize}

\subsection{Applications}
The proposed models in this work 
can be integrated into online platforms such as GitHub 
and help software teams automatically and instantly tag their issues with the correct label. One can also use the models dynamically to assign new priority labels. 
That is, teams can use the model periodically (e.g., at a specific hour each day) 
and re-assess the importance of issues based on the updated features 
(e.g., new discussions among team members, labels being added/removed, etc.).
Moreover, previous work has shown properly tagged issues are easier to manage. 
For instance, based on the determined objective (Bug report, Enhancement, support/Documentation), 
bug triaging can be facilitated and more important issues are assigned earlier to proper team members. 
Finally, major and important issue reports can be automatically selected 
to be included in software teams' periodic reports (such as release notes).

%%%%%%%%%%%%%%%%%%%%%%% Threats to Validity %%%%%%%%%%%%%%%%%%%%%%%%%%%%%%
%%%%%%%%%%%%%%%%%%%%%%%%%%%%%%%%%%%%%%%%%%%%%%%%%%%%%%%%%%%%%%%%%%%%%%

\subsection{Threats to Validity}\label{sec:threats}
In this section, we discuss the potential threats to the validity of our findings
and how we address or mitigate them.

\paragraph{Internal Validity}
Internal validity threats are related to our implementation and results,
labels analysis, and human bias in manual processes.
Although we have tried to thoroughly check our implementation, 
there still may be missed mistakes.
To mitigate this, we have made our code and data publicly available in our repository
for replication and use by other 
researchers.\footnote{\url{https://github.com/MalihehIzadi/IssueReportsManagement}}
Moreover, the parameters used in this study can pose potential threats.
To mitigate this we have tried to optimize all models 
and explicitly reported the values of parameters in each experiment separately.
Any unmentioned parameter is set to the default value of the corresponding library.
The set of synonym but differently written labels also poses a risk.
To mitigate this risk, two authors independently assessed these labels,
then compared the results, and resolved any case of conflict.
Moreover, in this process, both authors adhered to the labels' definitions provided by GitHub.
These measures increase our confidence in the manually created label sets.
% Moreover, we identify the clusters of issue labels manually.
% To minimize any potential bias, 
% the first author manually identified these clusters,
% while the second author verified them.
% As the definition of these labels are clear, 
% there was no conflict between the authors.

As the main goal for the human labeling and evaluation task 
is to showcase the ability of the model when prioritizing unlabeled issues, 
we were not able to compare against the ground truth labels in this experiment.
To mitigate this we employed a partially objective labeling task 
and took the majority vote for each label as its ground truth.
As prioritizing issues is a subjective task, biases and different opinions cannot be avoided. 
A factor that is important for an individual is not necessarily considered important for another person. 
Thus, the problem of prioritization is inherently subjective and biased.
We took several measures to mitigate such biases, including
selecting matured projects,
randomly selecting issues from these projects,
inviting a large number of professional developers and software engineers 
to participate in our experiment (diversity), 
providing labelers with important information of projects, 
and instructing them to get to know the project and its type of issues (awareness and knowledge).
We also assigned each labeler only to one project.
Moreover, we computed two inter-rater reliability measures, 
Percent Agreement and Randolph's free-margin multi-rater kappa.
The results indicated that there is a substantial level of agreement 
among labelers for all projects. 
Furthermore, for two projects there are perfect agreements. 
On one hand, the key limitation of Percent Agreement 
is that it does not account for the possibility that labelers may guess the labels,
so it may overestimate the true agreement~\mbox{\cite{mchugh2012interrater}}.
As our labelers are experts in the Software Engineering domain 
and are instructed to make well-informed decisions, 
little guessing is likely to exist, 
minimizing this risk.
On the other hand, due to some of the kappa's assumptions, 
it can underestimate the agreement among labelers~\mbox{\cite{mchugh2012interrater}}.
That is why we have included both of these measures in this study.
Moreover, participants' level of carefulness and effort 
can also affect the validity of experiment's results. 
To mitigate this risk, we recruited participants who
expressed interests in our research and double checked the
results to make sure there is no error.
For instance, we removed outlier labelers to avoid introducing noise 
by including people who had responded with low-quality labels 
(more than $50\%$ inconsistency with others).
In the end, it is worth mentioning, 
due to the inherent subjectiveness of the prioritization task,
the results of this particular experiment 
may not be completely replicable using another set of labelers or issue reports.

\paragraph{External Validity}
These threats are related to the generalizability of our work.
To address this issue, 
in both tasks we train our models on large-scale datasets.
For the objective-prediction task we use over $818K$ issue reports 
collected from approximately $61K$ repositories.
Furthermore, for the second, priority-detection, 
we also have trained a generic model in a cross-project context.
We have shown that our model can successfully predict 
priority of issue reports for unseen repositories.

\paragraph{Construct Validity}
Threats to construct validity relates to the suitability of the
evaluation metrics used in this study.
We use a set of standard evaluation metrics, 
namely Precision, Recall, F1-score, and Accuracy 
which are all employed in previous work~\cite{kallis2019ticket,song2020bee,huang2018automating} in the field.
However, more clusters and/or synonym labels can be found and used in the future.
Another threat is the choice of classifiers and the list of feature sets that we feed our models.
It is possible that using different features sets (and models)
result in different findings.
To address this issue, we thoroughly inspected issue reports and collected a large set of features.
Then we performed feature engineering methods to identify the most important ones.
We also used normalization techniques on numerical features.
Furthermore, we experimented with different Machine Learning models 
to find the best algorithm that fits our case.
To obtain more stable results for smaller datasets, 
we used the cross-validation technique.
However, random selection does not preserve chronology 
and ignores possible dependency between issue reports
that may have an influence on the trend of issues' category and importance in practice.
To mitigate this problem, the blocked version of cross-validation technique 
through adding margins can be used.

%%%%%%%%%%%%%%%%%%%%%%% Discussion %%%%%%%%%%%%%%%%%%%%%%%%%%%%%%%%%%%%%%%
%%%%%%%%%%%%%%%%%%%%%%%%%%%%%%%%%%%%%%%%%%%%%%%%%%%%%%%%%%%%%%%%%%%%%%

% \section{Discussion}\label{sec:discuss}
%  Our aim is to determine
% whether only categorical attributes (or only summary or text
% attributes) can train the classifier to achieve appropriate accuracy
% level for predicting bug priority of a new bug report
% or the features should be combined to achieve better results.

% Observation 2—Better performance does not imply
% more suitable for deployment. Recall that our classifiers will
% be deployed as a preliminary step to other software analytics
% modeling (e.g., defect prediction). For this reason, precision
% is a more important measure to optimize, since the cost of
% false positives is larger that of false negatives. Table II shows
% that rebalancing (SMOTE = Y) improves recall at the cost
% of precision. Even though the F1-scores tend to improve,
% indicating the benefit to recall is larger than the cost to
% precision, the importance of precision implies that the nonrebalanced
% classifiers may be more suitable for deployment.

% Replication Package. 
% All material and data of the bugs
% report used in our study as well as the developers’ anonymized
% answers are available in our replication package [18].

\section{Related Work}\label{sec:related}
In the following, we review studies related to
two phases of our proposed approach
in the categories of 
collective knowledge in SE,
issue report classification, 
issue report prioritization, 
and cross-project models.
% ----- Related ------
\subsection{Collective Knowledge in Software Engineering}
Collective knowledge accumulated on software-related platforms has been exploited in various work to help improve the software development process by introducing new techniques or providing empirical evidence.
Various types of collective knowledge have provided the means to perform studies on
empirical studies on such knowledge acquired from Stack Overflow, GitHub, and App stores
\cite{baltes2019sotorrent,wu2019developers,hu2019studying,zeng2019studying},
investigating, utilizing, and improving crowd-sourced knowledge in Stack Overflow
\cite{zhang2015survey,DBLPtavakoli},
usage of collective knowledge in a cross-platform setting
\cite{baltes2019usage}.
Vasilescu et al. \cite{vasilescu2013stackoverflow} 
studied the relationship between
Stack Overflow activities and the development process in GitHub through analyzing the available crowd-sourced knowledge. 
They claimed the Stack Overflow activity rate correlates with the code changing activity in GitHub. For instance, active committers tend to provide more answers on Q\&A websites.
In another work, Vasilescu et al. \cite{vasilescu2014social}
studied the evolution of mailing list participation
after the lunch of StackExchange.
They showed that the behavior of developers has been impacted by the emergence of these platforms, e.g., users are motivated to provide faster answers on StackExchange than on \texttt{r-help} mailing list due to its gamified environment.

There also numerous studies on providing automatic and intelligent solutions for various SE problems through exploiting these sources of collective knowledge such as
source code summarization 
\cite{wan2018improving,aghamohammadi2020generating},
automatic tag (topic) recommendation in Stack Overflow and GitHub
\cite{wang2018entagrec++,izadi2021topic} and more.
For instance, Zhou et al. \cite{zhou2020studying} 
through acknowledging the voluntary nature of open source software 
and the difficulty of finding appropriate developers 
to solve difficult yet important issue reports
studied monetary rewards (bounties) 
to motivate developers and help the evolution of the project.
They found the timing of bounties is the crucial factor 
affecting the likelihood of an issue being handled.
In another work, Chen et al. \cite{chen2021demystifying}
performed an empirical study on the user-provided logs in bug reports 
to investigate the problems that developers encounter 
and how to facilitate the diagnosis and fixing process.
Da Costa et al. \cite{da2018impact} conducted a comparative study on traditional and rapid release cycles to grasp the effect of rapid release cycles on the integration delay of fixed issues by analyzing 72K issue reports from the Firefox project.

Our work is similar to the above in the context that we too try to exploit collective knowledge to address SE problems and provide efficient and automatic solutions.
However, we specifically aim at addressing the management of issue reports as an important source of such knowledge to further facilitate and support the evolution of software projects.
We employ advanced Machine Learning techniques to address issue reports from two aspects of objective and priority and help team managers make better decisions. 
In the following, we review the literature on issue report classification and prioritization.

\subsection{Issue Report Classification}
Bug report categorization using Machine Learning techniques 
has received increasing attention from the software research community.
Antoniol et al. \cite{antoniol2008is} used three Machine Learning algorithms, 
namely Naive Bayes, Decision Trees and Logistic Regression
to determine whether the text
of their manually-labeled bug reports is enough to classify them into 
two distinctive classes of bugs or non-bugs. 
They found the information included in issue can be indeed used to classify them.
However, they only investigated three projects of 
Mozilla, Eclipse and JBoss projects from Bugzilla.
In the past years, there have been more researches on 
categorizing bug reports using text mining, 
topic modeling \cite{pingclasai2013classifying,limsettho2016unsupervised} 
and classification techniques \cite{sohrawardi2014comparative,zhou2016combining,terdchanakul2017bug,pandey2017automated}
in bug tracking systems such as Bugzilla and Jira .
% In the same field, Pingclasai et al. proposed a classification approach to identify bugs and non-bugs. They used Latent Dirichlet allocation (LDA) method with NB and LLR classifier. The precision of HTTPClient, Jackrabbit and Lucene projects varied from 66% to 76%, 65% to 77% and 71% to 82% respectively. The other work by Zhou et al. [16] in 2014, introduced the automatic bug prediction technique by combing text and data mining algorithms. The summary of bug reports are extracted by text mining algorithm and fed into the machine learner model with selected structured fields. The data drafting algorithm is used to combine the stages. The execution showed better results by increasing
% the f-measure. 

The main focus of the previous work has been on distinguishing
\textit{bug} from \textit{non-bug} reports for the purpose of bug triaging.
Moreover, most of these studies only investigate a few number of projects 
and rely only on a limited set of projects and their data 
for training separate models.
In fact, there is no proof whether 
they are suitable from a large-scale perspective.
Therefore, in the issue-objective prediction phase of our proposed approach, 
we perform a large-scale analysis of 
issue reports in GitHub issues 
to classify them into three coarse-grained classes of 
Bug, Enhancement, and Support 
using state-of-the-art transformer-based techniques.

To the best of our knowledge, there are two papers 
with the same focus (similar issue classes, GitHub as the common platform, large-scale) 
as the issue-objective prediction phase of our proposed approach.
In 2019, Kallis et al. \cite{kallis2019ticket} proposed \textit{TicketTagger}, 
a tool based on FastText for classifying issues to three categories of 
\texttt{Bug}, \texttt{Enhancement} and \texttt{Question}.
These categories are among the default labels of GitHub issue system.
They trained their model on the text (title and description)
of $30K$ issue reports from about $12K$ GitHub repositories.
Their evaluation reports $82\%$, $76\%$, and $78\%$ of Precision/Recall scores 
for three classes of Bug, Enhancement, and Question, respectively.
% The model is a multi-class linear neural model that receives
% the set of n-grams (i.e., sequences of n consecutive words)
% extracted from the issue title and description, and outputs the probability
% distribution of the issue over the predefined categories [30].
Recently, BEE was proposed by Song and Chaparro \cite{song2020bee}, 
which uses a the pre-trained model of TicketTagger to label issues. 
Then it proceeds to identify the structure of bug descriptions 
from predicted reports that are predicted to be a bug 
in the issue-objective prediction phase.
% In this paper, we make two novel contributions with respect to their prior work.
Furthermore, Huang et al.~\cite{huang2018automating} 
proposed a model based on convolutional neural networks (CNN) 
to classify issue report sentences.
They manually labeled $5408$ issue sentences from four GitHub projects 
and categorized them into seven groups based on their intentions.
These intention categories include 
Problem Discovery,
Feature Request,
Information Giving, 
Information Seeking,
Solution Proposal, 
Aspect Evaluation and finally, Meaningless.
Our objective categories overlap with theirs, 
however, they extract issue sentence intention while we classify the whole issue.
We not only classify issue reports and outperform these baselines, 
but also use these probabilities as one of the input features
in the second stage of our proposed approach 
which is the prioritization of said issues.
This is because the priority of issues is largely sensitive to their actual objective.
Although answering users' questions and helping them are important tasks, 
fixing bugs and adding new features are probably assigned higher ranks of importance.
Pull Requests are intertwined with issue reports 
(they usually try to address open issues),
and this notion is confirmed by Gousios et al.'s \cite{gousios2015work} findings as well.
They reported that software project integrators 
tend to prioritize contributions (pull requests) to their projects 
by considering the criticality of bug-fixes, 
the urgency of new features and their size.
Furthermore, to the best of our knowledge, 
we are the first to use transformer-based classifiers 
to manage issue reports.
We fine-tune RoBERTa, a pre-trained model on our large-scale dataset 
and achieve higher accuracy (outperforming all the baselines by large margins).
We also apply more rigorous text processing techniques 
and we employ a much larger dataset incorporating more labels
to train a more generic model.
More specifically, we collect and process 
over one million issue reports from $80K$ repositories.
We include $818K$ preprocessed issues 
from about $61K$ repositories, 
while Kallis et al.~\cite{kallis2019ticket} use only $30K$ issues from $12K$ repositories. 
% with only the three labels of bug, feature and question.

% ranking%%%%%%%%%%%%%%%%%%%%%%%%%%%%%%%%%%%%%%%%%%%%%%%%%%%
\subsection{Issue Report Prioritization}\label{related_pr}

Researchers have been studying 
bug report prioritization avidly~\cite{uddin2017survey}. 
Kanwal and Maqbool~\cite{kanwal2012bug} proposed a bug priority recommender
using SVM and NB classification techniques.
Alenezi and Banitaan~\cite{alenezi2013bug} tried to predict bug priority
using three classifiers, namely NB, DT, RF
on two projects of Firefox and Eclipse.
Tian et al.~\cite{tian2013drone} proposed \textit{DRONE} for Eclipse projects.
They investigated the effect of multiple factors, namely
temporal, textual, author, related-report, severity, and product features 
of bug reports on their priority level in the five-category ranking system 
of the Bugzilla's Bug Tracking System (BTS).
Kikas et al.~\cite{kikas2016using} proposed an approach to predict the lifetime of an
issue and whether it can be closed in a given period, 
using dynamic and contextual features of issues.
There are also other studies on 
prioritizing pull requests (PRioritizer) \cite{veen2015automatically},
and prioritizing user‐reported issues through their relations 
with app reviews and ratings~\cite{gao2015paid,noei2019too,noei2019towards,di2020investigating}.
For instance, Noei et al.~\cite{noei2019towards} 
proposed an approach to identify issues that need immediate attention 
through matching them with related user reviews in several apps.
They suggested software teams should first address 
issues that are mapped to the highest number of reviews.
By doing so, their app rating can be positively affected.

Although there are several studies on 
prioritizing bug reports on Bugzilla, 
the scope and features available in these systems differ.
For instance, Bugzilla is primarily designed for bug report management 
and has a predefined set of five priority labels, 
thus it has more training data available. 
In these studies, the focus is on bug reports and predicting whether a report is a bug or non-bug,
while we train a model to detect the objective behind opening issues.
It also includes information regarding the severity
which previous work has greatly exploited.
A recent study proposed by Dhasade et al.~\cite{dhasade2020towards}
has addressed the need 
for priority prediction models in GitHub.
However, they use LDA to identify the categories of issues,
then train a classifier to predict hotness of issue reports on a daily basis.
On contrary, our model uses classification models 
to label issues with two straightforward labels (High/Low). 
The model can be used on both open and closed issues. 
And it can be utilized both 
for prioritizing tasks plus resource allocation 
and also for documentation purposes such as writing reports,
delivering release notes, and highlighting the most important closed issues in a release.

The mentioned studies mostly try to prioritize bug reports 
in various ITS and BTS systems such as Bugzilla, Jira, and GitHub 
using different Machine Learning techniques. 
However, to the best of our knowledge,
there is no prior work on supervised models for issue reports of GitHub similar to ours.
Our work differs from bug prioritizing approaches 
since we address all issues and not just bugs.
Furthermore, BTS systems have readily available metadata 
which are unfortunately missing in GitHub.
Merten et al.~\cite{merten2016software} empirically analyzed 
four open-source projects from GitHub and Redmine
and found projects' metadata can improve classifier performance.
Therefore, we also conduct feature engineering techniques on metadata of issue reports 
and extract the salient features from GitHub which have not been utilized before.
On the other hand, approaches based on linking app's user reviews and issues from GitHub 
do not take into account various important factors 
such as author information, 
the amount of discussion happening in the report, 
issue lifetime, issue category, etc.
In addition to utilizing the metadata of reports,
we predict the objective of issue reports 
and then feed the predicted probability to our classification model for prioritizing.
We also perform sentiment analysis and include the outcome in our prioritizer model.
Moreover, we train both project-based and cross-project models. 
And finally, we also conducted a human labeling and evaluation task
to asses the performance of the proposed model 
on unseen data and provided developers' insights for future studies as well.
% ranking%%%%%%%%%%%%%%%%%%%%%%%%%%%%%%%%%%%%%%%%%%%%%%%%%%%
\subsection{Cross-project Models}
Peters et al.~\cite{peters2013better} 
claimed project-based predictors
are weak for small datasets. 
Also, Kitchenham et al.~\cite{kitchenham2007cross} found that
relying on project-based datasets is problematic 
due to the challenging task of 
collecting just enough project data 
to train models properly. 
Cross-project classification is a realistic solution 
for training a generic model 
from the data of a large number of different projects.
The trained model then can be successfully used for projects 
that have little to no data available for training.
In this field, a few studies have been conducted in the cross-project context.
Yu et al.~\cite{yu2018transferring} conducted an empirical study 
to identify the factors that affect the performances of transferring reusable
models across projects in the context of issue classification.
They extracted $28$ attributes grouped into four dimensions. 
Sharma et al.~\cite{sharma2012predicting} 
evaluated different Machine Learning models 
for predicting priority of new issues 
in five datasets of Open-Office and Eclipse projects.
Our work complements these studies 
by analyzing factors affecting issue reports in GitHub 
from both aspects of issue objective and priority.

%%%%%%%%%%%%%%%%%%%%%%%%%%%%%%%%%%%%%%%%%%%%%%%%%%%%%%%%%%%%%%%%%%%%%%
%%%%%%%%%%%%%%%%%%%%%%% Conclusion and Future Work %%%%%%%%%%%%%%%%%%%%%%%%%
%%%%%%%%%%%%%%%%%%%%%%%%%%%%%%%%%%%%%%%%%%%%%%%%%%%%%%%%%%%%%%%%%%%%%%

\section{Conclusions and Future Work}\label{sec:conclude}
An issue report can be opened 
due to several reasons including reporting bugs, 
requesting new features 
or merely for seeking support from the software team.
Naturally not all issues are equally important.
Some may require immediate care, 
some may need to be included 
in the documentation reports of the project,
while others are not as urgent.
In this study, we proposed a two-stage approach 
to predict both objective and importance of issue reports 
posted on software repositories.
We defined three sets of features related to issue reports
and exploited state-of-the-art text classifiers 
to achieve our goal.
According to the evaluation results,
our models outperform the baselines 
in both project-based and cross-project settings 
with $82\%$ and $75\%$ accuracy 
for objective and priority prediction, respectively.
Furthermore, we showed that our proposed priority prediction model 
in the cross-project setting performs on par with the project-based models.
Moreover, we conducted a human labeling and evaluation task 
to use the proposed priority detection model 
on unlabeled issue reports 
from six unseen projects with the help of $30$ Software Engineers. 
The results indicate that the model is capable of 
predicting priority of unseen data with high accuracy (90\%).
Therefore, our proposed model can be used 
for other unseen projects successfully 
without the need for extra training.

In the future, we plan to work on finer-grained categories 
of both objectives and priority levels.
Moreover, based on the results of our human labeling and evaluation experiment, 
we plan to investigate more features that 
can affect the importance of an issue from Software Engineers' perspectives.
For instance, many participants considered the \textit{bug type} and the degree of its \textit{impact} as an important factor while prioritizing issue reports.
Finally, adding more projects from other programming languages 
can also help generalizability of the proposed approach.

%%%%%%%%%%%%%%%%%%%%%%%%%%%%%%%%%%%%%%%%%%%%%%%%%%%%%%%%%%%%%%%%%%%%%%
%%%%%%%%%%%%%%%%%%%%%%% Appendix %%%%%%%%%%%%%%%%%%%%%%%%%
%%%%%%%%%%%%%%%%%%%%%%%%%%%%%%%%%%%%%%%%%%%%%%%%%%%%%%%%%%%%%%%%%%%%%%
\begin{appendices}
\section{Priority Labels}\label{appendix:priority_labls}
\autoref{tab:appendix_label_priority} presents the list of manually extracted labels from top GitHub repositories (most star) for categories of high and low priority issues.
\begin{table}[tb!]
\caption{Selected labels for each category of issue priority}
\label{tab:appendix_label_priority}
\begin{tabular}{p{20mm}p{90mm}}
\toprule
\textbf{Priority} & \textbf{Labels' list} \\
\toprule
High-priority 
& p0, priority: p0, p1, priority 1, priority: p1, priority 2, 
critical, criticalpriority, priority-critical,	critical priority,	priority:critical,	priority critical,	priority: critical,	priority - critical, critical-priority,	priority/critical,
urgent, priority/urgent, priority/blocker, priority: blocker, important, priority/important, 	
priority: major, highpriority,	priority-high,	high priority, priority:high, priority high,	priority: high,	priority - high, high-priority,	priority/high, is:priority	\\ \midrule
Low-priority	
& p3, priority: p3, priority 4, priority: minor, lowpriority, priority-low,	low priority,	priority:low,	priority low,	priority: low,	priority - low,	low-priority,	priority/low,	is:no-priority\\
\bottomrule
\end{tabular}
\end{table}

\end{appendices}

\bibliographystyle{spmpsci}
\bibliography{main}

\end{document}